# Performance Analysis of OTSM under Hardware Impairments and Imperfect CSI

Abed Doosti-Aref, *Member, IEEE*, Christos Masouros, *Fellow, IEEE*, Xu Zhu, *Senior Member, IEEE*, Ertugrul Basar, *Fellow, IEEE*, Sinem Coleri, *Fellow, IEEE*, and Huseyin Arslan, *Fellow, IEEE*

*Abstract*—Orthogonal time sequency multiplexing (OTSM) has been recently proposed as a single-carrier waveform offering similar bit error rate to orthogonal time frequency space (OTFS) and outperforms orthogonal frequency division multiplexing (OFDM) in doubly-spread channels (DSCs); however, with a much lower complexity making it a potential candidate for 6G wireless networks. In this paper, the performance of OTSM is explored by considering the joint effects of multiple hardware impairments (HWIs) such as in-phase and quadrature imbalance (IQI), direct current offset (DCO), phase noise, power amplifier non-linearity, carrier frequency offset, and synchronization timing offset for the first time in the area. First, the discrete-time baseband signal model is obtained in vector form under all mentioned HWIs. Second, the system input-output relations are derived in time, delay-time, and delay-sequency (DS) domains in which the parameters of all mentioned HWIs are incorporated. Third, analytical expressions are derived for the pairwise and average bit error probability under imperfect channel state information (CSI) as a function of the parameters of all mentioned HWIs. Analytical results demonstrate that under all mentioned HWIs, noise stays additive white Gaussian, effective channel matrix is sparse, DCO appears as a DC signal at the receiver interfering with only the zero sequency, and IQI redounds to self-conjugated sequency interference in the DS domain. Simulation results reveal the fact that by considering the joint effects of all mentioned HWIs and imperfect CSI not only OTSM outperforms OFDM by 29% in terms of energy of bit per noise but it performs same as OTFS in high mobility DSCs.

## I. INTRODUCTION

The sixth generation (6G) of wireless networks is characterized by the new technologies aiming at supporting higher reliability, energy efficiency, and spectral efficiency in high mobility vehicular networks. To accommodate the mentioned demands and bandwidth shortage, millimeter-wave (mmWave) and beyond frequency bands are being investigated to be employed in the 6G; however, hardware impairments (HWIs) are challenging issues in such frequency bands. Thus, performance evaluation of 6G candidate waveforms under HWIs is a crucial design step, which should be carefully explored for new waveform domains [1]-[4].

Abdollah (Abed) Doosti-Aref is with Sharif University of Technology a.doosti@sharif.edu, Christos Masouros is with University College London, c.masouros@ucl.ac.uk, Xu Zhu is with Harbin Institute of Technology, xuzhu@ieee.org, Ertugrul Basar and Sinem Coleri are with Koc University ebasar@ku.edu.tr, scoleri@ku.edu.tr, Huseyin Arslan is with Istanbul Medipol University, huseyinarslan@medipol.edu.tr.

### A. Motivations and Review of the Related Works

Orthogonal time sequency multiplexing (OTSM) has been recently proposed as a single-carrier (SC) waveform in delay-sequency (DS) domain [5], [6], which offers similar bit error rate (BER) to orthogonal time frequency space (OTFS) [7]; however, with a much lower complexity in doubly spread channels (DSCs) under high mobilities. Since OTSM is implemented based on Walsh-Hadamard transform (WHT), the modulation and demodulation in OTSM-based systems are merely done through addition/subtraction, which give rise to a lower implementation complexity in comparison with other waveforms implemented through multiplication [5]-[9]. Such attributes make OTSM transceivers as affordable equipment for SC communications in 6G wireless networks [4]-[6], [9].

Homodyne transceivers suit the implementation limitations in vehicular networks rather than heterodyne transceivers, which impose extra complexity and energy consumption to the system [1]. On the other hand, homodyne direct conversion transceivers suffer from multiple HWIs, such as in-phase and quadrature imbalance (IQI) [1], [10], [14]-[24], [26]-[28], direct current offset (DCO) [13], [15], [22], [23], and phase noise (PN) [1]-[3], [11], [12], happening at both transmitter (Tx) and receiver (Rx), power amplifier non-linearity (PAN) [16], [17], [20], occurring at the Tx along with carrier frequency offset (CFO) [1], [10]-[15], [23], [25], [28] and synchronization timing offset (STO) [12], [24], [25], betiding at the Rx. Such impairments are mainly due to the fabrication inaccuracies of analog components, which leads to performance degradation and energy inefficiency.

To evaluate the robustness of waveforms under HWIs and propose practical breakthroughs for HWI compensation (HWIC), obtaining the system input-output (I/O) relation by incorporating HWI parameters is a prerequisite step. This is because, the system I/O relation, the effective channel matrix, and noise properties alter under HWIs in time-frequency (TF) [10]-[15], delay-Doppler (DD) [2], [3], [18]-[25], DS [26]-[28], and delay-scale [29] waveform domains for narrowband and wideband applications. Knowing the specifications of noise and effective channel matrix in different waveform domains through the system I/O relation under HWIs enable one to design optimal receiver with the aim of HWIC. In [30], it is demonstrated that additive white Gaussian noise (AWGN) transforms to improper Gaussian noise in time domain due to Rx-IQI and optimum transceiver is accordingly designed for HWIC. In [18], [19], through the system I/O relation in DD domain, it is shown that in the presence of Tx/Rx-IQI, the transmitted signal experiences AWGN and a sparse effective channel, making the receiver design simpler in DD domain.

TABLE I
CONTRIBUTIONS OF THIS PAPER COMPARED TO THE STATE-OF-THE-ART

| Waveform, Domain | | OFDM, TF | | | | | OTFS, DD | | | | | OTSM, DS | | | |
|---|---|---|---|---|---|---|---|---|---|---|---|---|---|---|---|
| | Ref. | [10] | [11] | [12] | [13] | [15] | [3] | [18] | [20] | [23] | [24] | [26] | [27] | [28] | This Paper |
| Hardware Impairment | Tx/IQI | ✓ | × | × | × | × | × | ✓ | ✓ | ✓ | ✓ | × | ✓ | × | ✓ |
| | Rx/IQI | ✓ | × | × | × | ✓ | × | × | ✓ | ✓ | ✓ | ✓ | ✓ | ✓ | ✓ |
| | Tx/DCO | × | × | × | × | × | × | × | × | ✓ | × | × | × | × | ✓ |
| | Rx/DCO | × | × | × | ✓ | ✓ | × | × | × | ✓ | × | × | × | × | ✓ |
| | Tx/PN | × | × | × | × | × | ✓ | × | × | × | × | × | × | × | ✓ |
| | Rx/PN | × | ✓ | ✓ | × | × | × | × | × | × | × | × | × | × | ✓ |
| | PAN | × | × | × | × | × | × | × | ✓ | × | × | × | × | × | ✓ |
| | CFO | ✓ | ✓ | ✓ | ✓ | ✓ | × | × | × | ✓ | × | × | × | ✓ | ✓ |
| | STO | × | × | ✓ | × | × | × | × | × | × | ✓ | × | × | × | ✓ |
| System I/O Relation | | ✓ | ✓ | ✓ | ✓ | ✓ | ✓ | ✓ | ✓ | ✓ | ✓ | ✓ | ✓ | ✓ | ✓ |
| Imperfect CSI | | × | × | × | × | × | × | ✓ | × | × | × | × | × | × | ✓ |
| Analytical BER | | × | × | × | ✓ | × | × | ✓ | × | × | × | × | × | × | ✓ |

In addition, thanks to the system I/O relation under multiple HWIs in other waveform domains, we can take the advantages of AWGN and sparse effective channel, rather than time domain, to reduce the coupling effect, noise impact, and implementation complexity of joint HWIC by considering the correlation between HWIs in parallel multi task learning (MTL) based estimation algorithms [10].

From the HWIs point of view, OFDM and OTFS have been well studied in the literature [1]-[3], [10]-[25]. However, the number of works on HWI in OTSM-based systems is sporadic [26]-[28]. In [26], the system I/O relation is first obtained for OTSM transceiver by considering only IQI at Rx. Then, an embedded pilot arrangement is presented for joint HWIC and channel parameters estimation. In [27], the system I/O relation is derived for OTSM transceiver under only IQI at both Tx and Rx through which the training sequence is employed for joint HWIC and channel parameters estimation. In [28], the system I/O relation is obtained for OTSM transceiver by considering merely Rx-IQI and CFO. However, derivation of I/O relation and analytical study of BER by considering all HWIs listed in Table I under imperfect channel state information (CSI), which are fully explored in this paper for OTSM system, have not been yet addressed not only in the DS domain but also in other waveform domains.

*B. Contributions Compared to the State-of-the-Art*

Motivated by the above background, the contributions in this paper are precisely compared to the state-of-the-art in Table I, which are further detailed as follows.

i) We propose a discrete-time baseband signal model for OTSM homodyne transceiver in vector form through which the components of multiple HWIs are incorporated step by step to the signal and system models. In this context, for the first time in the literature, we consider jointly the memory effects of PAN at Tx, IQI, DCO, and PN at both Tx and Rx, along with CFO and STO at Rx. The proposed signal model can be utilized in vehicular networks for vehicle-to-vehicle, vehicle-to-base station, and high-speed train-to-vehicle links.

ii) The system I/O relations are obtained in time, delay-time (DT), and DS domains in both vector and matrix forms by incorporating the parameters of all HWIs listed in Table I for the first time in the area. In particular, it will be demonstrated that the I/O relations presented in [6], [26]-[28] are derived as the special cases from our proposed generalized model.

iii) Through analytical studies it is demonstrated that although the received noise is imposed by IQI, PN, and CFO at the Rx, its additive white Gaussian property is retained in the DS domain. Also, the effective channel matrix in the DS domain is sparse when all HWIs listed in Table I are jointly incorporated in the signal and system models. Moreover, Tx/Rx IQI in OTSM transceivers redounds to self-conjugated sequency interference (SCSI) of the channel impaired symbols along all their sequencies. However, Tx/Rx DCO appears as a zero-sequency signal at the Rx, introducing the zero sequency interference (ZSI) in the DS domain.

iv) Analytical results are presented for the upper bound of conditional pairwise error probability (PEP), unconditional PEP, and average bit error probability (ABEP) under imperfect CSI, as a function of the parameters of all HWIs listed in Table I for the first time in the literature.

v) Extensive simulation studies are carried out to evaluate the performance of OTSM by considering the joint effects of all HWIs listed in Table I under different scenarios for varying modulation orders, mobility speeds, and detectors for the first time in the literature. We further compare the performance of OTSM with orthogonal frequency division multiplexing (OFDM) and OTFS under HWIs as well as the ideal system including no HWIs under the same conditions. It should be noted that, to the best of the authors' knowledge, as listed in Table I, there is no work in the literature through which the performance of OTFS is evaluated by considering the joint effects of all HWIs listed in Table I. However, for the first time in the literature, this paper assesses the performance of OTFS system in which all HWIs are incorporated.

*C. Notations and organization of the paper*

Table II shows the list of abbreviations and notations used in this paper. The rest of the paper is organized as follows. The system model is described in Section II followed by the OTSM signal model under HWIs presented in Section III. Derivation of the system I/O relations in time, DT, and DS domains together with the associated remarks are explained in

TABLE II
LIST OF THE ABBREVIATIONS AND NOTATIONS

| Notation | Description |
|---|---|
| HWIs | Hardware Impairments |
| HWIC | Hardware Impairment Compensation |
| DSCs | Doubly Spread Channels |
| SC | Single-Carrier |
| IQI | In-phase and Quadrature Imbalance |
| DCO | Direct Current Offset |
| PN, CFO | Phase Noise, Carrier Frequency Offset |
| PAN | Power Amplifier Non-linearity |
| STO | Synchronization Timing Offset |
| SCSI | Self-Conjugated Sequency Interference |
| ZSI, DD | Zero Sequency Interference, Delay-Doppler |
| DS, DT | Delay-Sequency, Delay-Time |
| MDI | Mirror Doppler Interference |
| MTL | Multi Task Learning |
| PEP | Pairwise Error Probability |
| ABEP | Average Bit Error Probability |
| V2X | Vehicle to Everything |
| $\mathbf{A}$, $\mathbf{a}$ | Bold uppercase letter is matrix and bold lowercase letter is column vector |
| $\mathcal{CN}(\mu_0, \sigma_0^2)$ | Complex normal distribution with mean of $\mu_0$ and variance of $\sigma_0^2$ |
| $\mathcal{U}(a,b)$ | Uniform distribution in the interval [a,b] |
| $\mathbf{W}_N$, $\mathbf{I}_N$, $\mathbf{1}_N$, $\mathbf{0}_N$ | Normalized $N$-square WHT matrix, $N$-square identity matrix, all-ones, all-zeros column vectors of length $N$ |
| $(\cdot)^*$, $(\cdot)^T$, $(\cdot)^H$, $\|\cdot\|$ | Complex conjugate, transpose, conjugate transpose, and Frobenius norm operations |
| $\delta(\cdot)$, $\mathrm{E}(\cdot)$, $\Re(\cdot)$, $Q(\cdot)$ | Dirac delta, expectation, real-part, and Gaussian tail distribution functions |
| $\mathbb{R}^{M\times N}$, $\mathbb{C}^{M\times N}$ | The set of $M\times N$ dimensional matrices with real $\mathbb{R}$ and complex $\mathbb{C}$ entries |
| cov(**a**) | Covariance matrix of vector **a** |
| rank(**A**) | Rank of matrix **A** |
| trace(**A**) | Trace of matrix **A** |
| vec(**A**) | Column-wise vectorization of matrix **A** |
| $\mathrm{vec}_{M,N}^{-1}(\mathbf{a})$ | $M\times N$ matrix folded column-wise from the column vector **a** of length $NM$ |
| diag(**a**) | Transforms vector **a** into a diagonal matrix |
| $\otimes$, $\odot$, $\circledast$ | Kronecker product, Hadamard product, and dyadic convolution operations |

Section IV. Error performance analysis is described in Section V followed by the performance evaluation provided in Section VI. Finally, conclusions are summarized in Section VII.

## II. THE SYSTEM MODEL

We consider a vehicular network including vehicles moving at different speeds, base stations located at different places of the network, high-speed trains, and other fixed and mobile users as shown in Fig. 1. All vehicles are equipped with an OTSM-based homodyne transceiver, enabling communications of the vehicle with everything (V2X). For each vehicle a single antenna is adopted to emit the information symbols, which are statistically independent and identically distributed (i.i.d) with zero mean. The quadrature amplitude modulation (QAM) mapping is utilized to decorate $M\times N$ symbols in the DS domain as an OTSM frame, where $M$ and $N$ represent the number of delay and sequency bins of the DS grid, respectively. $N$ is considered to be a power of 2 and each OTSM frame occupies a bandwidth of $B = M\Delta f$ Hz and a time duration of $T_f = NT$ seconds, where $\Delta f$ and $T$ denote the sampling frequency interval and symbol time duration, respectively. In addition, we assume $T\Delta f = 1$, enabling critical sampling for any pulse shaping waveform [6].

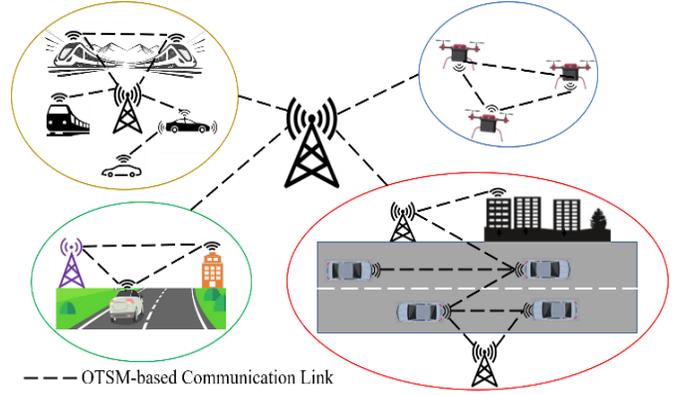

Fig. 1. Illustration of the OTSM links in vehicular network.

A time-varying DSC model with $P$ dominant reflectors is used to represent the channel for each V2X link. Accordingly, the channel impulse response (CIR) in the DD domain has a sparse configuration given as

$$h(\tau,\nu) = \sum_{i=1}^{P} h_i \delta(\tau - \tau_i)\delta(\nu - \nu_i), \quad (1)$$

where $h_i \sim \mathcal{CN}(0,1/P)$, $0 \leq \tau_i \leq \tau_{\max}$, and $-\nu_{\max} \leq \nu_i \leq \nu_{\max}$ denote the propagation gain, fractional delay shift, and fractional Doppler shift associated with the $i$th channel path, respectively. Also, $\ell_i = \tau_i B \in \mathbb{R}$ and $k_i = \nu_i T_f \in \mathbb{R}$ denote the normalized delay shift and normalized Doppler shift related to the $i$th channel path, respectively. Accordingly, $\ell_{\max}$ and $k_{\max}$ represent the maximum delay and Doppler shift indices of the channel, respectively. Hence, $\tau_{\max}$ and $\nu_{\max}$ represent the maximum delay and Doppler shift of the channel, which are constrained as $\tau_{\max}\nu_{\max} \ll 1$, i.e., under spread fading channel. Applying inverse fast Fourier transform to the Doppler dimension in (1), the CIR in DT domain is obtained as

$$g(\tau,t) = \int_{\nu} h(\tau,\nu) e^{j2\pi\nu(t-\tau)} d\nu = \sum_{i=1}^{P} h_i e^{j2\pi\nu_i(t-\tau_i)}. \quad (2)$$

Let $\mathcal{L} = \{0,...,l_{\max}\}$ denote the set of indices of delay bins, describing delay shifts at integer multiples of delay resolution $1/M\Delta f$. After applying sampling theorem to (2) at integer multiples of the delay resolution, namely $t = q/M\Delta f$ for $q = 0,...,NM + l_{\max} - 1$ and $\tau = l/M\Delta f$ for $l \in \mathcal{L}$, the discrete-time baseband CIR in the DT domain is given as [31]

$$g[l,q] = \sum_{i=1}^{P} h_i e^{j2\pi k_i (q-l)/MN} \mathrm{sinc}(l - \ell_i). \quad (3)$$

## III. THE OTSM SIGNAL MODEL UNDER HWIS

In this section, we obtain the signal model for the OTSM transceiver under IQI, DCO, PN, memory effects of PAN, CFO, and STO based on the block diagram depicted in Fig. 2.

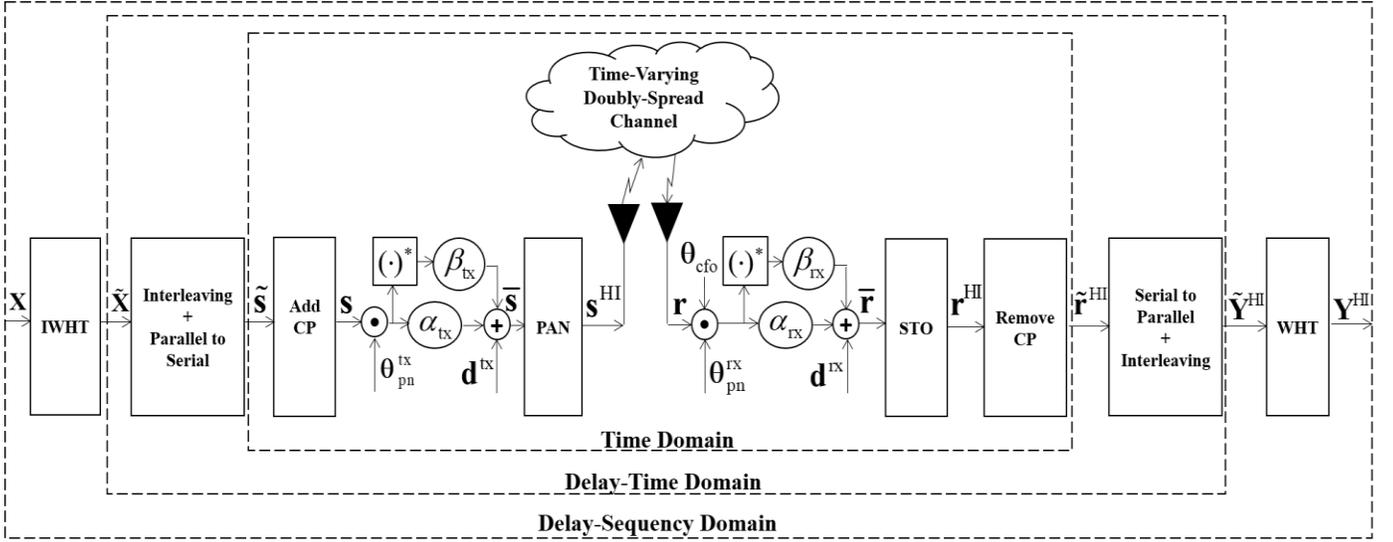

Fig. 2. The block diagram of OTSM transceiver in the presence of Tx/Rx HWIs.

*A. OTSM Transmitted Frame Before Incorporating Tx-HWIs*

Let $\mathbf{x} = [\mathbf{x}_0^T, ..., \mathbf{x}_{M-1}^T]^T \in \mathbb{C}^{NM \times 1}$ be the vector of information symbols in DS domain, where the symbol vectors $\mathbf{x}_m \in \mathbb{C}^{N \times 1}$ for $m = 0, ..., M-1$ correspond to the DS matrix representation of information symbols given as $\mathbf{X} = [\mathbf{x}_0, ..., \mathbf{x}_{M-1}]^T \in \mathbb{C}^{M \times N}$ after standing at each row of $\mathbf{X}$ to have $\mathbf{x} = \text{vec}(\mathbf{X}^T)$. Accordingly, the row and column indices of $\mathbf{X}$ correspond to the delay and sequency indices of the DS grid, respectively. To prevent inter block interference (IBI) and interference between the data and pilot, the last $2l_{\max}+1$ rows of $\mathbf{X}$ are set to zero, enabling interleaved zero padding (ZP) among the blocks [6]. Not only does this setting guard the tandem leakage but also facilitate the channel estimation in the time domain [6]. As depicted in Fig. 2, to move from the DS domain to DT domain, an $N$-point inverse WHT (IWHT) is applied along the sequency domain, namely each row of $\mathbf{X}$ as

$$\tilde{\mathbf{X}} = [\tilde{\mathbf{x}}_0, ..., \tilde{\mathbf{x}}_{M-1}]^T = \mathbf{X}\mathbf{W}_N. \quad (4)$$

Defining $\tilde{\mathbf{x}} = [\tilde{\mathbf{x}}_0^T, ..., \tilde{\mathbf{x}}_{M-1}^T]^T \in \mathbb{C}^{NM \times 1}$ as the input of DT domain, we have $\tilde{\mathbf{x}} = \text{vec}(\tilde{\mathbf{X}}^T)$. The DT samples are then column-wise vectorized to generate the time domain samples vector $\tilde{\mathbf{s}} = \text{vec}(\tilde{\mathbf{X}})$, which is further burst into $N$ blocks each of which of length $M$ in the time domain as $\tilde{\mathbf{s}} = [\tilde{\mathbf{s}}_0^T, ..., \tilde{\mathbf{s}}_{N-1}^T]^T \in \mathbb{C}^{NM \times 1}$. The vector representation of $\tilde{\mathbf{X}}$ gives rise to the interleaved transmission of the information symbols in the time domain, which can be readily represented in terms of $\tilde{\mathbf{x}}$ and the interleave perfect shuffle matrix $\mathbf{P}$ as

$$\tilde{\mathbf{s}} = \mathbf{P}\tilde{\mathbf{x}}. \quad (5)$$

Hence, due to the row-column interleaving, the $M$ samples of the time domain blocks $\tilde{\mathbf{s}}_n$ correspond to the DT symbol vectors $\tilde{\mathbf{x}}_m$ as $\tilde{\mathbf{s}}_n[m] = \tilde{\mathbf{x}}_m[n]$. This relation will be utilized in Subsection IV. B to obtain the I/O relation in the DT domain.

To represent $\tilde{\mathbf{s}}$ as the input of time domain channel in terms of $\mathbf{x}$ as the input of DS domain channel, we use the relation between the column-wise vectorization and Kronecker product in what follows. Considering the equation $\mathbf{V} = \mathbf{AXB}$, where $\mathbf{V}$, $\mathbf{A}$, $\mathbf{X}$, and $\mathbf{B}$ are given matrices, a more convenient form of the equation can be represented as

$$\text{vec}(\mathbf{V}) = (\mathbf{B}^T \otimes \mathbf{A})\text{vec}(\mathbf{X}). \quad (6)$$

When we apply this to (4) while using the relation between the perfect shuffle matrix $\mathbf{P}$ and column-wise vectorization for a given matrix $\mathbf{C}$, i.e., $\text{vec}(\mathbf{C}) = \mathbf{P}\text{vec}(\mathbf{C}^T)$, the transmitted vector in time domain in terms of the vector of information symbols in the DS domain can be written as

$$\tilde{\mathbf{s}} = (\mathbf{W}_N \otimes \mathbf{I}_M)\mathbf{P}\mathbf{x}. \quad (7)$$

Also, by using the property of the unitary permutation matrix $\mathbf{P}$ given as $(\mathbf{I}_M \otimes \mathbf{W}_N) = \mathbf{P}(\mathbf{W}_N \otimes \mathbf{I}_M)\mathbf{P}^T$, (7) can be written as

$$\tilde{\mathbf{s}} = \mathbf{P}(\mathbf{I}_M \otimes \mathbf{W}_N)\mathbf{x}. \quad (8)$$

To aid the channel estimation process, as depicted in Fig. 2, a cyclic prefix (CP) of length $l_{\max}$ is copied from the end of the time domain vector appended as a preamble at the head of the time domain vector given as

$$\mathbf{s} = \mathbf{A}_{\text{cp}}\tilde{\mathbf{s}} \in \mathbb{C}^{(NM+l_{\max}) \times 1}, \quad (9)$$

where $\mathbf{A}_{\text{cp}} = \begin{bmatrix} \mathbf{0}_{l_{\max} \times (NM-l_{\max})}, \mathbf{I}_{l_{\max}} \\ \mathbf{I}_{NM} \end{bmatrix} \in \mathbb{R}^{(NM+l_{\max}) \times NM}$ denotes the CP appending matrix.

*B. OTSM Transmitted Frame After Incorporating Tx-HWIs*

In the presence of IQI, DCO, and PN at the Tx, as depicted in Fig. 2, the time domain hardware impaired vector $\bar{\mathbf{s}}$ can be mathematically modeled as

$$\bar{\mathbf{s}} = \alpha_{\text{tx}}(\mathbf{s} \odot \boldsymbol{\theta}_{\text{pn}}^{\text{tx}}) + \beta_{\text{tx}}(\mathbf{s} \odot \boldsymbol{\theta}_{\text{pn}}^{\text{tx}})^* + \mathbf{d}^{\text{tx}}, \quad (10)$$

where $\mathbf{d}^{\text{tx}} = d^{\text{tx}}\mathbf{1}_{NM+l_{\max}} \in \mathbb{R}^{(NM+l_{\max}) \times 1}$ denotes the DCO at Tx. We assume the practical asymmetrical model through which the Tx-IQI coefficients are given as $\alpha_{\text{tx}} = (1 + g_{\text{tx}}e^{-j\varphi_{\text{tx}}/2})/2$ and $\beta_{\text{tx}} = (1 - g_{\text{tx}}e^{j\varphi_{\text{tx}}/2})/2$, where $g_{\text{tx}}$ and $\varphi_{\text{tx}}$ represent the gain and phase mismatches, respectively. In addition, the oscillator output at Tx with carrier frequency of $f_c$ and PN of $\theta_{pn}^{tx}(t)$ is modeled as $e^{-j(2\pi f_c t + \theta_{pn}^{tx}(t))}$ [1]. Accordingly, the Tx-PN sampled at $t = q/M\Delta f$ in the discrete-time equivalent baseband model can be given as $S_{pn}^{tx}[q] = e^{-j\theta_{pn}^{tx}(q/M\Delta f)}$ for $q = 0,...,NM + l_{max} - 1$. Consequently, $\boldsymbol{\theta}_{pn}^{tx}$ in (10) can be written as

$$\boldsymbol{\theta}_{pn}^{tx} = [S_{pn}^{tx}[0],...,S_{pn}^{tx}[NM + l_{max} - 1]]^T \in \mathbb{C}^{(NM + l_{max}) \times 1}. \quad (11)$$

To include PAN and memory effects in the signal vector at Tx, we assume memory polynomial based on the complexity-reduced Volterra model, which includes memory effects and exponentiated envelope terms [16], [17], [20]. Accordingly, by considering $\bar{\mathbf{s}}[q]$ as the input of power amplifier (PA), as shown in Fig. 2, the discrete baseband samples of hardware impaired vector $\mathbf{s}^{HI}$ at the output of PA under Tx-IQI, Tx-DCO, Tx-PN, and PAN are obtained by

$$\mathbf{s}^{HI}[q] = \sum_{j=0}^{N_\rho - i} \sum_{i=0}^{M_\rho} \rho_{ij} \bar{\mathbf{s}}[q-i] |\bar{\mathbf{s}}[q-i]|^j, \quad (12)$$

where the parameters $\rho_{ij}$, $M_\rho$, and $N_\rho$ are the linearization coefficients, non-linear memory depth, and non-linearity order, respectively [16], [17].

Finally, after pulse shaping of time domain samples $\mathbf{s}^{HI}[q]$ followed by digital to analog converter, the hardware impaired analog signal $s^{HI}(t)$ is transmitted to the time-varying DSC.

*C. OTSM Received Frame Before Incorporating Rx-HWIs*
The channel impaired signal at the Rx is represented as $r(t) = \int_0^{\tau_{max}} g(\tau,t) s^{HI}(t-\tau) d\tau + w(t)$, where the additive noise $w(t)$ is distributed as $\mathcal{CN}(0, \sigma_0^2)$. Sampling $r(t)$ with a time period of $T_s = 1/M\Delta f$ at $t = qT_s$ for $q = 0,...,NM + l_{max} - 1$, the discrete-time baseband channel impaired signal in the time domain can be readily written as

$$\mathbf{r}[q] = \sum_{l \in \mathcal{L}} g[l,q] \mathbf{s}^{HI}[q-l] + \mathbf{w}[q], \quad (13)$$

Where $\mathbf{w}$ is a Gaussian vector with $\text{cov}(\mathbf{w}) = \sigma_0^2 \mathbf{I}_{NM + l_{max}}$.

*D. OTSM Received Frame After Incorporating IQI, DCO, PN, and CFO at the Rx*
In the presence of IQI, DCO, PN, and CFO at the Rx, as depicted in Fig. 2, the time domain channel impaired vector $\bar{\mathbf{r}}$ can be mathematically modeled as

$$\bar{\mathbf{r}} = \alpha_{rx}(\mathbf{r} \odot \boldsymbol{\theta}_{pn}^{rx} \odot \boldsymbol{\theta}_{cfo}) + \beta_{rx}(\mathbf{r} \odot \boldsymbol{\theta}_{pn}^{rx} \odot \boldsymbol{\theta}_{cfo})^* + \mathbf{d}^{rx}, \quad (14)$$

where $\mathbf{d}^{rx} = d^{rx} \mathbf{1}_{NM + l_{max}} \in \mathbb{R}^{(NM + l_{max}) \times 1}$ represents the Rx-DCO vector. Also $\alpha_{rx} = (1 + g_{rx} e^{-j\varphi_{rx}/2})/2$ and $\beta_{rx} = (1 - g_{rx} e^{j\varphi_{rx}/2})/2$ stand for the asymmetrical Rx-IQI coefficients in which $g_{rx}$ and $\varphi_{rx}$ represent the gain and phase imbalances at Rx, respectively. Here, we have assumed that the oscillator output at Rx with carrier frequency of $f_c$, CFO of $f_o$, and PN of $\theta_{pn}^{rx}(t)$ is modeled as $e^{-j(2\pi(f_c + f_o)t + \theta_{pn}^{rx}(t))}$ [1]. Thus, the Rx-PN and CFO stochastic processes in the discrete-time baseband model sampled at $t = q/M\Delta f$ for $q = 0,...,NM + l_{max} - 1$ can be represented as $S_{pn}^{rx}[q] = e^{-j\theta_{pn}^{rx}(q/M\Delta f)}$ and $S_{cfo}[q] = e^{-j(2\pi f_o q/M\Delta f)}$, respectively. Therefore, the Rx-PN vector $\boldsymbol{\theta}_{pn}^{rx}$ and CFO phase vector $\boldsymbol{\theta}_{cfo}$ in (14) can be written as

$$\boldsymbol{\theta}_{pn}^{rx} = [S_{pn}^{rx}[0],...,S_{pn}^{rx}[NM + l_{max} - 1]]^T \in \mathbb{C}^{(NM + l_{max}) \times 1}, \quad (15)$$

$$\boldsymbol{\theta}_{cfo} = [S_{cfo}[0],...,S_{cfo}[NM + l_{max} - 1]]^T \in \mathbb{C}^{(NM + l_{max}) \times 1}. \quad (16)$$

*E. OTSM Received Frame After Incorporating the STO*
To include the STO in $\bar{\mathbf{r}}$, we utilize the Gardner's timing synchronization loop at Rx through which the optimal symbol timing instants are estimated [24]. Accordingly, the samples of time domain hardware impaired received vector $\mathbf{r}^{HI}$ under Rx-IQI, Rx-DCO, Rx-PN, CFO, and STO can be written as

$$\mathbf{r}^{HI}[q] = \sum_{i=I_1}^{I_2} \bar{\mathbf{r}}[m_q - i] h_{intp}[i + \mu_q], \quad (17)$$

where $I_1$ and $I_2$ are finite integers and $m_q$, $\mu_q$, and $h_{intp}$ are the base-point for the $q$th interpolant, fractional interval, and impulse response of the interpolating filter, respectively.

*F. OTSM Received Frame in the DT and DS Domains*
Let $\mathbf{R}_{cp} = [\mathbf{0}_{NM \times l_{max}}, \mathbf{I}_{NM}] \in \mathbb{R}^{NM \times (NM + l_{max})}$ denote the CP removal matrix as depicted in Fig. 2. Then, the CP is removed from the time domain vector as

$$\tilde{\mathbf{r}}^{HI} = \mathbf{R}_{cp} \mathbf{r}^{HI} \in \mathbb{C}^{NM \times 1}. \quad (18)$$

As shown in Fig. 2, after removing the CP, the time domain channel impaired vector $\tilde{\mathbf{r}}^{HI}$ is folded back into a matrix to show the DT matrix as $\tilde{\mathbf{Y}}^{HI} = \text{vec}_{M,N}^{-1}(\tilde{\mathbf{r}}^{HI}) = [\tilde{\mathbf{y}}_0^{HI},...,\tilde{\mathbf{y}}_{M-1}^{HI}]^T \in \mathbb{C}^{M \times N}$. Defining $\tilde{\mathbf{y}}^{HI} = [(\tilde{\mathbf{y}}_0^{HI})^T,...,(\tilde{\mathbf{y}}_{M-1}^{HI})^T]^T$ as the output vector of DT domain, the relation between $\tilde{\mathbf{r}}^{HI}$, $\tilde{\mathbf{Y}}^{HI}$, and $\tilde{\mathbf{y}}^{HI}$ is given as

$$\tilde{\mathbf{r}}^{HI} = \text{vec}(\tilde{\mathbf{Y}}^{HI}) = \mathbf{P} \tilde{\mathbf{y}}^{HI}. \quad (19)$$

Therefore, due to the column-wise folding, the $M$ samples of blocks $\tilde{\mathbf{r}}_n^{HI}$ are related to the DT symbol vectors $\tilde{\mathbf{y}}_m^{HI}$ as $\tilde{\mathbf{r}}_n^{HI}[m] = \tilde{\mathbf{y}}_m^{HI}[n]$. This relation will be utilized in Subsection IV. B to obtain the system I/O relation in the DT domain.

To move from the DT domain to DS domain, as depicted in Fig. 2, $N$-point WHT is applied along the time dimension, namely each row of $\tilde{\mathbf{Y}}^{HI}$ as

$$\mathbf{Y}^{HI} = [\mathbf{y}_0^{HI},...,\mathbf{y}_{M-1}^{HI}]^T = \tilde{\mathbf{Y}}^{HI} \mathbf{W}_N. \quad (20)$$

Considering $\mathbf{y}^{HI} = [(\mathbf{y}_0^{HI})^T,...,(\mathbf{y}_{M-1}^{HI})^T]^T \in \mathbb{C}^{NM \times 1}$ as the output vector of DS domain at Rx, the relation between $\mathbf{Y}^{HI}$ and $\mathbf{y}^{HI}$ in terms of the perfect shuffle matrix $\mathbf{P}$ is written as $\mathbf{P} \mathbf{y}^{HI} = \text{vec}(\mathbf{Y}^{HI})$. Applying (6) to (20) and then by using $\tilde{\mathbf{r}}^{HI} = \text{vec}(\tilde{\mathbf{Y}}^{HI})$, $(\mathbf{I}_M \otimes \mathbf{W}_N) = \mathbf{P}(\mathbf{W}_N \otimes \mathbf{I}_M) \mathbf{P}^T$, and $\mathbf{P} \mathbf{y}^{HI} = \text{vec}(\mathbf{Y}^{HI})$, the vector of the received symbols at the output of DS domain under HWIs can be finally written as

$$\mathbf{y}^{HI} = (\mathbf{I}_M \otimes \mathbf{W}_N) \mathbf{P}^T \tilde{\mathbf{r}}^{HI}. \quad (21)$$

## IV. THE SYSTEM INPUT-OUTPUT RELATIONS UNDER HWIS

In this section, we obtain the system I/O relations in time, DT, and DS domains under joint effects of HWIs listed in Table I.

### A. Vector-Form I/O Relation in the Time Domain

To represent the I/O relation in terms of the delay index $m$ and the sequency or time index $n$ of the OTSM frame in the DS or DT domains, respectively, the discrete-time index $q$ in time domain can be separated into $m$ and $n$ as $q = m + nM$, for $m = 0,...,M-1$ and $n = 0,...,N-1$. Accordingly, we define $g_1[l, m+nM]$, $g_2[l, m+nM]$, $g_3[l, m+nM]$, and $\tilde{w}_n[m]$ as

$$g_1[l, m+nM] = (\alpha_{tx}\alpha_{rx}S_{pn}^{tx}[m+nM]S_{pn}^{rx}[m+nM]$$
$$S_{cfo}[m+nM]g[l,m+nM] + \beta_{tx}^*\beta_{rx}S_{pn}^{tx}[m+nM] \quad (22)$$
$$S_{pn}^{rx*}[m+nM]S_{cfo}[m+nM]g^*[l,m+nM]),$$

$$g_2[l,m+nM] = (\beta_{tx}\alpha_{rx}S_{pn}^{tx*}[m+nM]S_{pn}^{rx}[m+nM]$$
$$S_{cfo}[m+nM]g[l,m+nM] + \alpha_{tx}^*\beta_{rx}S_{pn}^{tx*}[m+nM] \quad (23)$$
$$S_{pn}^{rx*}[m+nM]S_{cfo}^*[m+nM]g^*[l,m+nM]),$$

$$g_3[l,m+nM] = (\alpha_{rx}S_{pn}^{rx}[m+nM]S_{cfo}[m+nM]g[l,m+nM] + \quad (24)$$
$$\beta_{rx}S_{pn}^{rx*}[m+nM]S_{cfo}^*[m+nM]g^*[l,m+nM]),$$

$$\tilde{w}_n[m] = (\alpha_{rx}S_{pn}^{rx}[m+nM]S_{cfo}[m+nM]\tilde{w}[m+nM] + \quad (25)$$
$$\beta_{rx}S_{pn}^{rx*}[m+nM]S_{cfo}^*[m+nM]\tilde{w}^*[m+nM]),$$

and by putting (13) in (14), then (14) in (17), and after removing the CP, the I/O relation in time domain is given as

$$\tilde{r}_n^{HI}[m] = \sum_{l \in \mathcal{L}} g_1[l,m+nM]\tilde{s}_n[m-l] + g_2[l,m+nM] \quad (26)$$
$$\tilde{s}_n^*[m-l] + g_3[l,m+nM]\tilde{d}_n^{tx}[m] + \tilde{d}_n^{rx}[m] + \tilde{w}_n[m].$$

It is worth mentioning that in (26) we have assumed that $\tilde{\mathbf{d}}^{rx} = \mathbf{R}_{cp}\mathbf{d}^{rx} = d^{rx}\mathbf{1}_{NM}$ and $\tilde{\mathbf{d}}^{tx} = d^{tx}\mathbf{1}_{NM}$ to obtain the I/O relation for the input of the system before inserting the CP and the output of the system after removing the CP in time domain as shown in Fig. 2.

### B. Vector-Form I/O Relation in the DT Domain

By replacing $\tilde{r}_n^{HI}[m]$ with $\tilde{\mathbf{y}}_m^{HI}[n]$, $\tilde{s}_n[m]$ with $\tilde{\mathbf{x}}_m[n]$, and $g_i[l,m+nM]$ with $\tilde{\mathbf{g}}_{i_{m,l}}[n]$ for $i = 1,2,3$ in (26), the system I/O relation in the DT domain is obtained as

$$\tilde{\mathbf{y}}_m^{HI}[n] = \sum_{l \in \mathcal{L}} \tilde{\mathbf{g}}_{1_{m,l}}[n]\tilde{\mathbf{x}}_{m-1}[n] + \tilde{\mathbf{g}}_{2_{m,l}}[n]\tilde{\mathbf{x}}_{m-1}^*[n] + \tilde{\mathbf{g}}_{3_{m,l}}[n]\tilde{\mathbf{d}}_m^{tx}[n] + \quad (27)$$
$$\tilde{\mathbf{d}}_m^{rx}[n] + \tilde{\mathbf{w}}_m[n].$$

### C. Vector-Form I/O Relation in Terms of the Sequency Spread Matrix in the DS Domain

To obtain the system I/O relation in the DS domain, we apply $N$-point WHT along the time dimension of the DT received vector in (27), i.e., each row of the DT matrix $\tilde{\mathbf{Y}}^{HI}$ obtained through (27) for $n = 0,...,N-1$. Then, by utilizing the multiplication property of WHT, i.e., $\mathbf{W}_N(\mathbf{a} \odot \mathbf{b}) = (\mathbf{W}_N\mathbf{a}) \boxasterisk (\mathbf{W}_N\mathbf{b})$, for two arbitrary given vectors $\mathbf{a}$ and $\mathbf{b}$ of length $N$, the system I/O relation in the DS domain is obtained as

$$\mathbf{y}_m^{HI} = \sum_{l \in \mathcal{I}} \tilde{\mathbf{u}}_{1_{m,l}} \boxasterisk \mathbf{x}_{m-1} + \tilde{\mathbf{u}}_{2_{m,l}} \boxasterisk \mathbf{x}_{m-1}^* + \tilde{\mathbf{u}}_{3_{m,l}} \boxasterisk \hat{\mathbf{d}}_m^{tx} + \hat{\mathbf{d}}_m^{rx} + \hat{\mathbf{w}}_m, \quad (28)$$

where $\tilde{\mathbf{u}}_{i_{m,l}} = \mathbf{W}_N \tilde{\mathbf{g}}_{i_{m,l}}$ for $i = 1,2,3$ are the sequency spread vectors, $\mathbf{x}_{m-1} = \mathbf{W}_N \tilde{\mathbf{x}}_{m-1}$, $\hat{\mathbf{w}}_m = (\mathbf{W}_N \tilde{\mathbf{w}}_m)$, $\hat{\mathbf{d}}_m^{tx} = (\mathbf{W}_N \tilde{\mathbf{d}}_m^{tx}) = [Nd^{tx}, \mathbf{0}_{N-1}^T]^T$, and $\hat{\mathbf{d}}_m^{rx} = (\mathbf{W}_N \tilde{\mathbf{d}}_m^{rx}) = [Nd^{rx}, \mathbf{0}_{N-1}^T]^T$. Let $\mathbf{U}_{i_{m,l}} = \mathbf{W}_N \tilde{\mathbf{G}}_{i_{m,l}} \mathbf{W}_N$ denote the sequency spread matrices where for $i = 1,2,3$ we have $\tilde{\mathbf{G}}_{i_{m,l}} = \text{diag}(\tilde{\mathbf{g}}_{i_{m,l}}[0],...,\tilde{\mathbf{g}}_{i_{m,l}}[N-1])$. Accordingly, the system I/O relation in (28) can be rewritten as

$$\mathbf{y}_m^{HI} = \sum_{l \in \mathcal{I}} \mathbf{U}_{1_{m,l}} \mathbf{x}_{m-1} + \mathbf{U}_{2_{m,l}} \mathbf{x}_{m-1}^* + \mathbf{U}_{3_{m,l}} \hat{\mathbf{d}}_m^{tx} + \hat{\mathbf{d}}_m^{rx} + \hat{\mathbf{w}}_m. \quad (29)$$

The I/O relation given in (29) is further shown in Fig. 3 for $N = M = 8$ and $l_{max} = 1$. Since we have $\hat{\mathbf{d}}_0^{tx} = \hat{\mathbf{d}}_1^{tx} = ... = \hat{\mathbf{d}}_{M-1}^{tx}$ and $\hat{\mathbf{d}}_0^{rx} = \hat{\mathbf{d}}_1^{rx} = ... = \hat{\mathbf{d}}_{M-1}^{rx}$, the multiplication terms $\mathbf{U}_{3_{m,l}} \hat{\mathbf{d}}_m^{tx} = \mathbf{U}_{3_{m,l}} \hat{\mathbf{d}}_i^{tx}$ for $i = 0,...,M-1$ have been used for the sake of presentation in the third term given in Fig. 3. In other words, each row of the $NM \times NM$ dimensional matrix in the third term can be written in different forms by considering the mentioned property; however, for all representations of the matrix, its bandwidth, i.e., the maximum number of non-zero entries in each of its rows, becomes $N(l_{max}+1)$.

### D. Matrix-Form I/O Relations in Terms of the Effective Channel Matrices in the Time, DT, and DS Domains

Because of the ZP symbols used in DS domain, the IBI is removed [6] and as a result the I/O relation in time domain in terms of the transmitted hardware impaired samples $\mathbf{s}^{HI}$ and banded channel matrix $\mathbf{H} \in \mathbb{C}^{(NM+l_{max}) \times (NM+l_{max})}$ is given as

$$\mathbf{r} = \mathbf{Hs}^{HI} + \mathbf{w}. \quad (30)$$

Accordingly, the I/O relation in time domain for the input of the system before CP insertion namely $\tilde{\mathbf{s}}$ and the output of the system after CP removal namely $\tilde{\mathbf{r}}^{HI}$, as depicted in Fig. 2, in terms of the effective channel matrix $\mathbf{G} \in \mathbb{C}^{NM \times NM}$ and noise vector $\tilde{\mathbf{w}}$ is represented as

$$\tilde{\mathbf{r}}^{HI} = \mathbf{G}\tilde{\mathbf{s}} + \tilde{\mathbf{w}}, \quad (31)$$

where $\mathbf{G} = \mathbf{R}_{cp}\mathbf{HA}_{cp}$.

Substituting (5) and (19) in (31), the system I/O relation in the DT domain is obtained as

$$\tilde{\mathbf{y}}^{HI} = \tilde{\mathbf{G}}\tilde{\mathbf{x}} + \tilde{\mathbf{w}}, \quad (32)$$

where $\tilde{\mathbf{G}} = \mathbf{P}^T \mathbf{R}_{cp} \mathbf{HA}_{cp} \mathbf{P}$ is the block-diagonal effective channel matrix in the DT domain, i.e., every block of samples can be independently analyzed in the DT domain.

To obtain the I/O relation in DS domain, 8 substitutions are done from (8) to (21) as follows. First, (8) is substituted in (9), then (9) in (10), then (10) in (12), then (12) in (13), then (13) in (14), then (14) in (17), then (17) in (18), and finally by putting (18) in (21), the matrix form of the system I/O relation in the DS domain is obtained as (33), where, $\boldsymbol{\Theta}_{pn}^{tx} = \text{diag}(\boldsymbol{\theta}_{pn}^{tx})$, $\boldsymbol{\Theta}_{pn}^{rx} = \text{diag}(\boldsymbol{\theta}_{pn}^{rx})$, and $\boldsymbol{\Theta}_{cfo} = \text{diag}(\boldsymbol{\theta}_{cfo})$ denote the Tx-PN, Rx-PN, and CFO phase matrices, respectively.

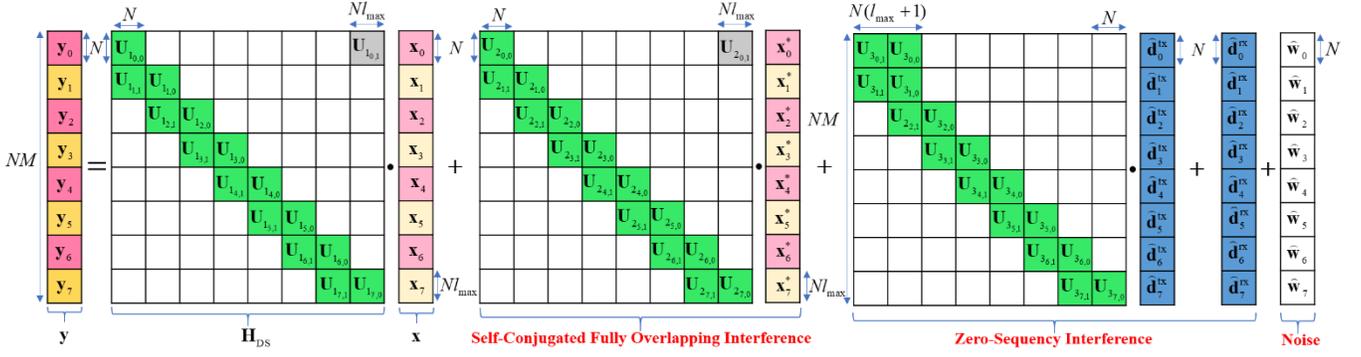

Fig. 3. Illustration of the system I/O relation in the DS domain for OTSM system under HWIs, $N=M=8$, $l_{\max}=1$.

$$\mathbf{y}^{\mathrm{HI}} = \underbrace{[(\mathbf{I}_M \otimes \mathbf{W}_N)\mathbf{P}^{\mathrm{T}}(\alpha_{\mathrm{tx}}\alpha_{\mathrm{rx}}\mathbf{R}_{\mathrm{cp}}\mathbf{\Theta}_{\mathrm{pn}}^{\mathrm{rx}}\mathbf{\Theta}_{\mathrm{cfo}}\mathbf{H}\mathbf{\Theta}_{\mathrm{pn}}^{\mathrm{tx}}\mathbf{A}_{\mathrm{cp}} + \beta_{\mathrm{tx}}^*\beta_{\mathrm{rx}}\mathbf{R}_{\mathrm{cp}}\mathbf{\Theta}_{\mathrm{pn}}^{\mathrm{rx}\,*}\mathbf{\Theta}_{\mathrm{cfo}}^*\mathbf{H}^*\mathbf{\Theta}_{\mathrm{pn}}^{\mathrm{tx}}\mathbf{A}_{\mathrm{cp}})\mathbf{P}(\mathbf{I}_M \otimes \mathbf{W}_N)]}_{\mathbf{H}_{\mathrm{DS}}:\ \text{Delay-sequency effective channel matrix}}\mathbf{x}$$
<div align="center">Signal term</div>

$$+ \underbrace{[(\mathbf{I}_M \otimes \mathbf{W}_N)\mathbf{P}^{\mathrm{T}}(\beta_{\mathrm{tx}}\alpha_{\mathrm{rx}}\mathbf{R}_{\mathrm{cp}}\mathbf{\Theta}_{\mathrm{pn}}^{\mathrm{rx}}\mathbf{\Theta}_{\mathrm{cfo}}\mathbf{H}\mathbf{\Theta}_{\mathrm{pn}}^{\mathrm{tx}\,*}\mathbf{A}_{\mathrm{cp}} + \alpha_{\mathrm{tx}}^*\beta_{\mathrm{rx}}\mathbf{R}_{\mathrm{cp}}\mathbf{\Theta}_{\mathrm{pn}}^{\mathrm{rx}\,*}\mathbf{\Theta}_{\mathrm{cfo}}^*\mathbf{H}^*\mathbf{\Theta}_{\mathrm{pn}}^{\mathrm{tx}\,*}\mathbf{A}_{\mathrm{cp}})\mathbf{P}(\mathbf{I}_M \otimes \mathbf{W}_N)]}_{\text{Conjugated signal term}}\mathbf{x}^*$$
<div align="center">Self-interference term: Self-conjugated fully-overlapping sequency interference</div>

$$+ \underbrace{[(\mathbf{I}_M \otimes \mathbf{W}_N)\mathbf{P}^{\mathrm{T}}(\alpha_{\mathrm{rx}}\mathbf{R}_{\mathrm{cp}}\mathbf{\Theta}_{\mathrm{pn}}^{\mathrm{rx}}\mathbf{\Theta}_{\mathrm{cfo}}\mathbf{H}\mathbf{A}_{\mathrm{cp}} + \beta_{\mathrm{rx}}\mathbf{R}_{\mathrm{cp}}\mathbf{\Theta}_{\mathrm{pn}}^{\mathrm{rx}\,*}\mathbf{\Theta}_{\mathrm{cfo}}^*\mathbf{H}^*\mathbf{A}_{\mathrm{cp}})]d^{\mathrm{tx}}\mathbf{1}_{NM}}_{\text{DC due to Tx-DCO imposed by channel, Rx PN, and CFO}} + \underbrace{[(\mathbf{I}_M \otimes \mathbf{W}_N)\mathbf{P}^{\mathrm{T}}]d^{\mathrm{rx}}\mathbf{1}_{NM}}_{\text{DC due to Rx-DCO}}$$
<div align="center">Self-interference term: Zero-sequency interference</div>

$$+ \underbrace{[(\mathbf{I}_M \otimes \mathbf{W}_N)\mathbf{P}^{\mathrm{T}}(\alpha_{\mathrm{rx}}\mathbf{R}_{\mathrm{cp}}\mathbf{\Theta}_{\mathrm{pn}}^{\mathrm{rx}}\mathbf{\Theta}_{\mathrm{cfo}}\mathbf{w} + \beta_{\mathrm{rx}}\mathbf{R}_{\mathrm{cp}}\mathbf{\Theta}_{\mathrm{pn}}^{\mathrm{rx}\,*}\mathbf{\Theta}_{\mathrm{cfo}}^*\mathbf{w}^*)]}_{\text{Noise term: Imposed by Rx-IQI, Rx-PN, and CFO}}. \qquad (33)$$

*E. Some Remarks on the Effective Channel Matrix, Noise, Sequency Interference, and I/O Relations in the DS Domain*

The I/O relation in (33) gives useful insights and information in terms of the DS channel matrix, interference, noise, and the state of the received signal to interference and noise. Accordingly, some remarks are discussed in what follows.

*Remark 1 (Effective Channel Matrix Is Sparse in the DS Domain under HWIs):* In the first term in (33), the DS channel matrix under HWIs, named as $\mathbf{H}_{\mathrm{DS}}$, is a diagonal matrix whose bandwidth is equal to $N(l_{\max}+1)$. Thus, the maximum number of non-zero entries over the number of all entries in $\mathbf{H}_{\mathrm{DS}}$ is equal to $N(l_{\max}+1)/(NM+l_{\max})^2$, which can be quite well approximated as $((l_{\max}+1)/NM^2) \ll 1$ since $NM \gg l_{\max}$, revealing the fact that the effective channel matrix in the DS domain has a sparse representation.

*Remark 2 (Tx/Rx IQI Redounds to Self-Conjugated Fully-Overlapping Sequency Interference in the DS Domain):* The second term in (33) represents the self-interference in terms of the conjugated symbols in DS domain. In analogy to the self-degradation and mirror Doppler interference (MDI) introduced along the Doppler axis in the DD domain in the IQI-OTFS systems [18], the channel impaired symbols in the DS domain deal with sequency interference in IQI-OTSM. As depicted in Fig. 4, the sequency values of the channel impaired symbols and intercurrent conjugated symbols are the same. Therefore, in contrary to IQI-OTFS in which we have MDI, i.e., intercurrent conjugated symbols concurrently appear at the symmetric Doppler taps $-k_i$ and $k_i$, these symbols interfere with the channel impaired symbols fully overlapping at the same sequency bins in IQI-OTSM over all delay bins, which may lead to the degradation of signal power. In other words, IQI in OTSM transceivers redounds to self-conjugated sequency interference (SCSI) of the symbols along all their sequencies in the DS domain as illustrated in Fig. 4.

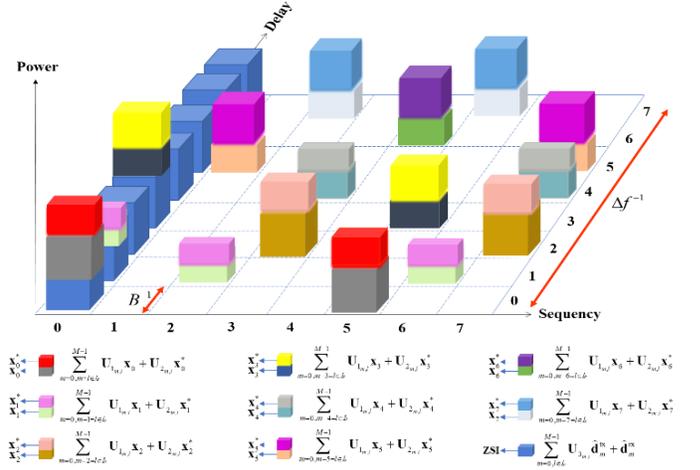

Fig. 4. Illustration of the SCSI and ZSI in OTSM frame in the presence of Tx/Rx IQI and Tx/Rx DCO in the DS domain.

*Remark 3 (Tx/Rx DCO Appears as a DC Signal Interfering with Only the Zero Sequency of Symbols in the DS Domain):* The third and fourth terms in (33) correspond to the Tx-DCO and Rx-DCO, respectively, which appear as DC signals in the

DS domain. As depicted in Fig. 4, Tx/Rx DCO interferes with only the zero sequency of the channel impaired symbols, which is named zero sequency interference (ZSI).

*Remark 4 (Noise Remains White Gaussian in the DS Domain under HWIs):* The fifth term in (33), which is also given as $\hat{\mathbf{w}}$ in the vector form of (29), describes the noise vector in the DS domain whose covariance matrix is given as

$$\text{cov}(\hat{\mathbf{w}}) = (1+\cos(2\omega) + g_{\text{rx}}^2(1-\cos(2\omega+\varphi_{\text{rx}})))\sigma_0^2/2\mathbf{I}_{NM}, \quad (34)$$

where, $\omega = 2\pi f_o q + \theta_{\text{pn}}^{\text{rx}}(q/M\Delta f)$ for $q = 0,...,NM-1$ corresponds to the CFO and Rx-PN; please refer to Appendix A to see the proof. Accordingly, we can conclude Theorem 1 as follows.

*Theorem 1:* Noise is statistically white Gaussian in the DS domain under Rx-IQI, Rx-DCO, Rx-PN, CFO, and STO.

*Proof:* Please refer to Appendix A to see the proof.

According to Theorem 1, in the absence of Rx-IQI, Rx-PN, and CFO namely $\varphi_{\text{rx}} = 0°$, $g_{\text{rx}} = 1$, $\theta_{\text{pn}}^{\text{rx}}(t) = 0$, $f_o = 0$, the noise amplification is ideally equal to zero in the DS domain. Also, from (33) and (34), we conclude that CFO and Rx-PN have no influence on the noise in the absence of Rx-IQI; however, as represented in (34), Rx-IQI amplifies the noise variance $(1+g_{\text{rx}}^2(1-\cos(\varphi_{\text{rx}}))/2)$ times even in the absence of Rx-PN and CFO. Therefore, under the condition that we have no CFO and Rx-PN, the amplified noise can be roughly ignored if the gain and phase imbalances are negligible; as a result, the noise in time and DS domains become identically distributed.

*Remark 5 (Obtaining the System I/O Relations Given in [6], [26]-[28] as the Special Cases of Our Proposed System I/O Relations):* Considering an ideal system with no HWIs, i.e., $g_{\text{tx}} = g_{\text{rx}} = 1$, $\varphi_{\text{tx}} = \varphi_{\text{rx}} = 0°$, $\theta_{\text{pn}}^{\text{tx}}(t) = \theta_{\text{pn}}^{\text{rx}}(t) = d^{\text{tx}} = d^{\text{rx}} = f_o = 0$, which results in $\alpha_{\text{tx}} = \alpha_{\text{rx}} = 1$, $\beta_{\text{tx}} = \beta_{\text{rx}} = 0$, $\boldsymbol{\Theta}_{\text{pn}}^{\text{tx}} = \boldsymbol{\Theta}_{\text{pn}}^{\text{rx}} = \boldsymbol{\Theta}_{\text{cfo}} = \mathbf{I}_{NM+l_{\max}}$, the system I/O relation in (33) is simplified to (35), which was first obtained in [6].

$$\mathbf{y} = (\mathbf{I}_M \otimes \mathbf{W}_N)\mathbf{P}^T\mathbf{G}\mathbf{P}(\mathbf{I}_M \otimes \mathbf{W}_N)\mathbf{x} + (\mathbf{I}_M \otimes \mathbf{W}_N)\mathbf{P}^T\mathbf{R}_{\text{cp}}\mathbf{w}. \quad (35)$$

Considering only Rx-IQI and no other HWIs, i.e., $\varphi_{\text{tx}} = 0°$, $g_{\text{tx}} = 1$, $d^{\text{tx}} = d^{\text{rx}} = 0$, and $\theta_{\text{pn}}^{\text{tx}}(t) = \theta_{\text{pn}}^{\text{rx}}(t) = f_o = 0$, which results in $\alpha_{\text{tx}} = 1$, $\beta_{\text{tx}} = 0$, and $\boldsymbol{\Theta}_{\text{pn}}^{\text{tx}} = \boldsymbol{\Theta}_{\text{pn}}^{\text{rx}} = \boldsymbol{\Theta}_{\text{cfo}} = \mathbf{I}_{NM+l_{\max}}$, the system I/O relation in (33) is simplified to (36), which was first obtained in [26].

$$\mathbf{y}^{\text{HI}} = [\alpha_{\text{rx}}(\mathbf{I}_M \otimes \mathbf{W}_N)\mathbf{P}^T\mathbf{R}_{\text{cp}}\mathbf{H}\mathbf{A}_{\text{cp}}\mathbf{P}(\mathbf{I}_M \otimes \mathbf{W}_N)]\mathbf{x} + $$
$$[\beta_{\text{rx}}(\mathbf{I}_M \otimes \mathbf{W}_N)\mathbf{P}^T\mathbf{R}_{\text{cp}}\mathbf{H}\mathbf{A}_{\text{cp}}\mathbf{P}(\mathbf{I}_M \otimes \mathbf{W}_N)]\mathbf{x}^* + \quad (36)$$
$$[(\mathbf{I}_M \otimes \mathbf{W}_N)(\alpha_{\text{rx}}\mathbf{P}^T\mathbf{R}_{\text{cp}}\mathbf{w} + \beta_{\text{rx}}\mathbf{P}^T\mathbf{R}_{\text{cp}}\mathbf{w}^*].$$

Considering the joint Tx/Rx IQI with no other HWIs, i.e., $d^{\text{tx}} = d^{\text{rx}} = 0$, which results in $\tilde{\mathbf{d}}^{\text{tx}} = \tilde{\mathbf{d}}^{\text{rx}} = \mathbf{0}_{NM}$, the system I/O relation in terms of the sequency spread matrices in (29) is simplified to (37), which was first obtained in [27].

$$\mathbf{y}_m^{\text{HI}} = \sum_{l \in L_0} \mathbf{U}_{1_{m,l}} \mathbf{x}_{m-l} + \mathbf{U}_{2_{m,l}} \mathbf{x}_{m-l}^* + \hat{\mathbf{w}}_m. \quad (37)$$

Considering joint Rx-IQI and CFO with no other HWIs, i.e., $\varphi_{\text{tx}} = 0°$, $g_{\text{tx}} = 1$, $d^{\text{tx}} = d^{\text{rx}} = 0$, and $\theta_{\text{pn}}^{\text{tx}}(t) = \theta_{\text{pn}}^{\text{rx}}(t) = 0$, which results in $\alpha_{\text{tx}} = 1$, $\beta_{\text{tx}} = 0$, $S_{\text{pn}}^{\text{tx}}[m+nM] = 1$, and $S_{\text{pn}}^{\text{rx}}[m+nM] = 1$, the system I/O relation in (27) is simplified to (38), which was first obtained in [28].

$$\tilde{\mathbf{y}}_m^{\text{HI}}[n] = \sum_{l \in L_0} \tilde{\mathbf{g}}_{1_{m,l}}[n]\tilde{\mathbf{x}}_{m-l}[n] + \tilde{\mathbf{g}}_{2_{m,l}}[n]\tilde{\mathbf{x}}_{m-l}^*[n] + \tilde{\mathbf{w}}_m[n]. \quad (38)$$

## V. ERROR PERFORMANCE ANALYSIS

In this section, thanks to the system I/O relation obtained in (33) in the DS domain and by assuming the optimal maximum likelihood (ML) criterion, we obtain the ABEP in the presence of all HWIs listed in Table I and imperfect CSI.

To ease the derivation of ABEP, we first represent the time domain channel matrix $\mathbf{H} \in \mathbb{C}^{(NM+l_{\max}) \times (NM+l_{\max})}$ as $\mathbf{H} = \sum_{i=1}^{P} h_i \boldsymbol{\Pi}^{\ell_i} \boldsymbol{\Delta}^{k_i}$ where $h_i \sim \mathcal{CN}(0, 1/P)$ is the gain of the $i$th channel path and the matrices $\boldsymbol{\Pi}$ and $\boldsymbol{\Delta}$ model the delay and Doppler shifts, respectively. $\boldsymbol{\Pi}^{\ell_i}$ can be obtained through $\ell_i$ times forward cyclic shift to the rows of $\mathbf{I}_{NM+l_{\max}}$ and $\boldsymbol{\Delta}^{k_i} = \text{diag}(z^0,...,z^{NM+l_{\max}})$ where $z = e^{j2\pi k_i/NM}$. Accordingly, we can rewrite the I/O relation in the DS domain given in (33) as

$$\mathbf{y}^{\text{HI}} = \boldsymbol{\Omega}_1(\mathbf{x})\mathbf{h} + \boldsymbol{\Omega}_2(\mathbf{x})\mathbf{h}^* + \overline{\mathbf{w}}, \quad (39)$$

where for $i = 1, 2,...,P$ we have

$$\boldsymbol{\Omega}_1(\mathbf{x}) = [\boldsymbol{\Omega}_{1_1}(\mathbf{x})\ \boldsymbol{\Omega}_{1_2}(\mathbf{x})\ ...\ \boldsymbol{\Omega}_{1_P}(\mathbf{x})] \in \mathbb{C}^{NM \times P}, \quad (40)$$

$$\boldsymbol{\Omega}_2(\mathbf{x}) = [\boldsymbol{\Omega}_{2_1}(\mathbf{x})\ \boldsymbol{\Omega}_{2_2}(\mathbf{x})\ ...\ \boldsymbol{\Omega}_{2_P}(\mathbf{x})] \in \mathbb{C}^{NM \times P}, \quad (41)$$

$$\boldsymbol{\Omega}_{1_i}(\mathbf{x}) = (\mathbf{I}_M \otimes \mathbf{W}_N)\mathbf{P}^T[\alpha_{\text{tx}}\alpha_{\text{rx}}\mathbf{R}_{\text{cp}}\boldsymbol{\Theta}_{\text{pn}}^{\text{rx}}\boldsymbol{\Theta}_{\text{cfo}}\boldsymbol{\Pi}^{\ell_i}\boldsymbol{\Delta}^{k_i}\boldsymbol{\Theta}_{\text{pn}}^{\text{tx}}$$
$$\mathbf{A}_{\text{cp}}\mathbf{P}(\mathbf{I}_M \otimes \mathbf{W}_N)\mathbf{x} + \beta_{\text{tx}}\alpha_{\text{rx}}\mathbf{R}_{\text{cp}}\boldsymbol{\Theta}_{\text{pn}}^{\text{rx}}\boldsymbol{\Theta}_{\text{cfo}}\boldsymbol{\Pi}^{\ell_i}\boldsymbol{\Delta}^{k_i}\boldsymbol{\Theta}_{\text{pn}}^{\text{tx}*}\mathbf{A}_{\text{cp}} \quad (42)$$
$$\mathbf{P}(\mathbf{I}_M \otimes \mathbf{W}_N)\mathbf{x}^* + \alpha_{\text{rx}}\mathbf{R}_{\text{cp}}\boldsymbol{\Theta}_{\text{pn}}^{\text{rx}}\boldsymbol{\Theta}_{\text{cfo}}\boldsymbol{\Pi}^{\ell_i}\boldsymbol{\Delta}^{k_i}\mathbf{A}_{\text{cp}}d^{\text{tx}}\mathbf{1}_{NM}],$$

$$\boldsymbol{\Omega}_{2_i}(\mathbf{x}) = (\mathbf{I}_M \otimes \mathbf{W}_N)\mathbf{P}^T[\beta_{\text{tx}}^*\beta_{\text{rx}}\mathbf{R}_{\text{cp}}\boldsymbol{\Theta}_{\text{pn}}^{\text{rx}*}\boldsymbol{\Theta}_{\text{cfo}}^*\boldsymbol{\Pi}^{\ell_i*}\boldsymbol{\Delta}^{k_i*}\boldsymbol{\Theta}_{\text{pn}}^{\text{tx}}$$
$$\mathbf{A}_{\text{cp}}\mathbf{P}(\mathbf{I}_M \otimes \mathbf{W}_N)\mathbf{x} + \alpha_{\text{tx}}^*\beta_{\text{rx}}\mathbf{R}_{\text{cp}}\boldsymbol{\Theta}_{\text{pn}}^{\text{rx}*}\boldsymbol{\Theta}_{\text{cfo}}^*\boldsymbol{\Pi}^{\ell_i*}\boldsymbol{\Delta}^{k_i*}\boldsymbol{\Theta}_{\text{pn}}^{\text{tx}*}\mathbf{A}_{\text{cp}} \quad (43)$$
$$\mathbf{P}(\mathbf{I}_M \otimes \mathbf{W}_N)\mathbf{x}^* + \beta_{\text{rx}}\mathbf{R}_{\text{cp}}\boldsymbol{\Theta}_{\text{pn}}^{\text{rx}*}\boldsymbol{\Theta}_{\text{cfo}}^*\boldsymbol{\Pi}^{\ell_i*}\boldsymbol{\Delta}^{k_i*}\mathbf{A}_{\text{cp}}d^{\text{tx}}\mathbf{1}_{NM}],$$

$$\mathbf{h} = [h_1, h_2,...,h_P] \in \mathbb{C}^{P \times 1}, \quad (44)$$

$$\overline{\mathbf{w}} = \hat{\mathbf{w}} + (\mathbf{I}_M \otimes \mathbf{W}_N)\mathbf{P}^T d^{\text{rx}}\mathbf{1}_{NM}. \quad (45)$$

Since $\hat{\mathbf{w}}$ has been shown through Theorem 1 as a Gaussian vector with zero mean and the covariance matrix given in (34), and the second term in (45) is a constant vector, $\overline{\mathbf{w}}$ becomes Gaussian with the mean vector of $d^{\text{rx}}\mathbf{1}_{NM}$ and covariance matrix of $\sigma_{\overline{\mathbf{w}}}^2 \mathbf{I}_{NM}$ where

$$\sigma_{\overline{\mathbf{w}}}^2 = ((1+\cos(2\omega) + g_{\text{rx}}^2(1-\cos(2\omega+\varphi_{\text{rx}})))\sigma_0^2/2 + d^{\text{rx}^2}). \quad (46)$$

Accordingly, the ML criterion can be written as

$$\hat{\mathbf{x}}^{\text{ML}} = \arg\min_{\mathbf{x} \in \mathbb{C}^{NM}} \|\mathbf{y}^{\text{HI}} - \boldsymbol{\Omega}_1(\mathbf{x})\mathbf{h} - \boldsymbol{\Omega}_2(\mathbf{x})\mathbf{h}^*\|^2. \quad (47)$$

To practically consider the limitations of channel estimation, we define the imperfect CSI vector $\overline{\mathbf{h}}$ in terms of perfect CSI $\mathbf{h}$, estimation error vector $\boldsymbol{\varepsilon} \sim \mathcal{CN}(0,1)$, and $b \in [0,1]$ as [32]

$$\overline{\mathbf{h}} = \sqrt{1-b^2}\mathbf{h} + b\boldsymbol{\varepsilon}, \quad (48)$$

where $b = 0$ leads to the results for the case of perfect CSI hereinafter. Using (47) and (48), we first obtain a tight upper bound for conditional PEP, defined as the probability of transmitting $x_i$ and erroneously detecting $x_j$ conditioned on imperfect CSI vector $\overline{\mathbf{h}}$ through Theorem 2 as follows.

*Theorem 2:* Considering $\boldsymbol{\Omega}(\mathbf{x}) = \boldsymbol{\Omega}_1(\mathbf{x}) + \boldsymbol{\Omega}_2(\mathbf{x})$ as the equivalent codeword matrix, the upper bound of conditional PEP under HWIs based on the ML detection can be written as

$$\Pr(x_i \to x_j | \bar{\mathbf{h}}) \leq Q\left(\sqrt{\frac{\|(\boldsymbol{\Omega}(x_i) - \boldsymbol{\Omega}(x_j))\bar{\mathbf{h}}\|^2}{2b^2 \|\boldsymbol{\Omega}(x_i)\|^2 + 2\sigma_{\bar{\mathbf{w}}}^2}}\right), \quad (49)$$

where $\boldsymbol{\Omega}(x_i) \in \mathbb{C}^{1 \times P}$ represents the $i$th row of matrix $\boldsymbol{\Omega}(\mathbf{x})$.

*Proof:* Please refer to Appendix B to see the proof.

Because of the high computation complexity of (49) in practice, to obtain a tractable form for PEP we then approximate the tail distribution function in (49) quite well by $Q(x) \leq e^{-x^2/2}/12 + e^{-2x^2/3}/4$. Accordingly, the upper bound of PEP can be written as [33]

$$\Pr(x_i \to x_j) \leq$$
$$E_{\bar{\mathbf{h}}}\left(\frac{1}{12}e^{-\rho_1 \|(\boldsymbol{\Omega}(x_i) - \boldsymbol{\Omega}(x_j))\bar{\mathbf{h}}\|^2} + \frac{1}{4}e^{-\rho_2 \|(\boldsymbol{\Omega}(x_i) - \boldsymbol{\Omega}(x_j))\bar{\mathbf{h}}\|^2}\right), \quad (50)$$

where $\rho_1 = 1/(4b^2 \|\boldsymbol{\Omega}(x_i)\|^2 + 4\sigma_{\bar{\mathbf{w}}}^2)$ and $\rho_2 = 1/(3b^2 \|\boldsymbol{\Omega}(x_i)\|^2 + 3\sigma_{\bar{\mathbf{w}}}^2)$. In addition, the probability density function of Gaussian random vector $\bar{\mathbf{h}}$ is given as [34]

$$f(\bar{\mathbf{h}}) = \frac{\pi^{-\kappa_1}}{\det(\boldsymbol{\Upsilon}_P)} e^{(-\bar{\mathbf{h}}^H \boldsymbol{\Upsilon}_P^{-1} \bar{\mathbf{h}})}, \quad (51)$$

where $\boldsymbol{\Upsilon}_P = E(\bar{\mathbf{h}}\bar{\mathbf{h}}^H) \in \mathbb{C}^{P \times P}$ and $1 \leq \kappa_1 = \text{rank}(\boldsymbol{\Upsilon}_P) \leq P$. It is worth mentioning that $\boldsymbol{\Upsilon}_P$ is a submatrix centered along the main diagonal of the covariance matrix of channel coefficients vector of length $NM$ when the DS effective channel matrix is captured at time instants $t = q/M\Delta f$ for $q = 0,...,NM-1$. Thus, due to the ZP used among the blocks and no IBI during the detection, $\boldsymbol{\Upsilon}_P$ is valid for all subblocks. Hence, by using (50), (51), and the spectral theorem [35], [36] for calculating the expectation in (50), the upper bound of PEP is given as

$$\Pr(x_i \to x_j) \leq \frac{1/12}{\det(\mathbf{I}_P + \rho_1 \boldsymbol{\Upsilon}_P \boldsymbol{\Phi})} + \frac{1/4}{\det(\mathbf{I}_P + \rho_2 \boldsymbol{\Upsilon}_P \boldsymbol{\Phi})}, \quad (52)$$

where $\boldsymbol{\Phi} = (\boldsymbol{\Omega}(x_i) - \boldsymbol{\Omega}(x_j))^H (\boldsymbol{\Omega}(x_i) - \boldsymbol{\Omega}(x_j))$.

Let $\kappa_2 = \text{rank}(\boldsymbol{\Phi})$ and $\lambda_i$ denote the non-zero eigen values of $\boldsymbol{\Phi}$. Since $h_i \sim \mathcal{CN}(0, 1/P)$, the non-zero eigen values of $\boldsymbol{\Upsilon}_P$ are all equal to $1/P$. Accordingly, (52) can be written as

$$\Pr(x_i \to x_j) \leq \frac{1}{12} \prod_{i=1}^{\kappa} \frac{1}{1 + \frac{\rho_1 \lambda_i}{P}} + \frac{1}{4} \prod_{i=1}^{\kappa} \frac{1}{1 + \frac{\rho_2 \lambda_i}{P}}, \quad (53)$$

where $\kappa = \text{rank}(\boldsymbol{\Upsilon}_P \boldsymbol{\Phi})$. In the high signal-to-noise ratio (SNR) regime, (53) can be readily approximated as

$$\text{PEP}^{\text{High-SNR}} \cong (12(\frac{\rho_1}{P})^\kappa \prod_{i=1}^\kappa \lambda_i)^{-1} + (4(\frac{\rho_2}{P})^\kappa \prod_{i=1}^\kappa \lambda_i)^{-1}. \quad (54)$$

*Remark 6 (Diversity Order of the OTSM System under Joint Effects of Multiple HWIs and Imperfect CSI):* From (54), it can be concluded that the diversity order of the OTSM system is determined by $\kappa$, which is upper bounded by $\kappa \leq \min\{\kappa_1, \kappa_2\}$ according to the rank inequality [35]. Also, when the receiver makes a single decision error out of $Q$-ary symbols, we have $\min\{\kappa_2\} = 1$. Thus, $\kappa$ can take values from the interval $[1, P]$.

Eventually, by applying the union bounding technique to (53), the ABEP of OTSM system under HWIs and imperfect CSI can be evaluated by

$$\text{ABEP} \leq \frac{1}{N_b 2^{N_b}} \sum_{x_i} \sum_{x_j} \Pr(x_i \to x_j) N_{be}(x_i, x_j), \quad (55)$$

where $N_b = N(M - 2l_{\max} - 1) \log_2^Q$ denotes the number of bits within each OTSM frame and $N_{be}$ represents the number bit errors when $x_i$ is erroneously detected as $x_j$.

## VI. PERFORMANCE EVALUATION

In this section, we assess the performance of OTSM system by considering the joint effects of all HWIs listed in Table I. Monte Carlo simulations are carried out through MATLAB to compare the numerical results with the analytical derivations presented in the previous sections. We assume a DS grid including $M$=64 delay bins and $N$=64 sequency bins to transmit each OTSM frame through a carrier frequency of 40 GHz over a bandwidth of 10 MHz with an excess pilot power of 3 dB while the imperfectness variance in the CSI vector is considered to be $b^2 = 0.001$. Information symbols in each OTSM frame are encoded with a rate of 1/2 through the low-density parity-check (LDPC) encoder with codeword length of 672 while the ML detector is employed unless otherwise is stated. The power delay profile vector of [0, -1.5, -1.4, -3.6, -0.6, -9.1, -7, -12, -16.9] in dB and tap delay vector of [0, 30, 150, 310, 370, 710, 1090, 1730, 2510] in ns associated with $P$=9 paths are utilized based on the standard Extended Vehicular A (EVA) channel model [37]. The index of delay shifts follows a uniform distribution of $\mathcal{U}(0, l_{\max})$ and the Doppler shifts are generated based on the Jakes's spectrum, i.e., $k_i = k_{\max} \cos(\xi_i)$ where $\xi_i \sim \mathcal{U}(0, 2\pi)$ over the $i$th path with fractional values of delay and Doppler. The HWI parameter settings are given in Table III. It is worth mentioning that 4 scenarios are defined in Table III to include the practical range of variation for the values of HWI parameters. The values are set in a fashion that the intensity of HWIs increase from Scenario 1 to Scenario 4. Accordingly, in the following simulation examples, Scenario 0 describes the ideal system experiencing no HWIs with perfect CSI and Scenario 4 represents the conditions under which the system undergoes the harshest HWIs at both Tx and Rx.

### A. Performance Evaluation at Fixed SNR per Bit

In this simulation example, we evaluate the average BER for 4-QAM OTSM system in terms of each HWI by considering the joint effects of all HWIs listed in Table III. The results are obtained at fixed value of signal-to-noise ratio per bit $E_b/N_0$, which will be henceforth denoted as SNR.

Figs. 5(a) to 5(i) show the analytical and numerical results of BER as a function of HWI parameters listed in Table III at two fixed SNR values of 20 dB and 23 dB. It should be noted that each sub-figure of Fig. 5 illustrates the BER directly in terms of a hardware impairment while the parameters of other HWIs have been fixed at their harshest values listed in Table III as Scenario 4. Figs. 5(a) to 5(d) show the BER performance in terms of the Tx-HWIs and Figs 5(e) to 5(i) illustrate it with respect to Rx-HWIs. In all sub-figures, it is observed that the larger the HWI values, the worse the BER performance becomes; however, the numerical and analytical results are perfectly the same when the BER is evaluated in terms of the HWI parameters at fixed SNR value.

TABLE III
HWI PARAMETER SETTINGS

| HWIs | Tx-IQI | Tx-DCO | Tx-PN | PAN | Rx-IQI | Rx-DCO | Rx-PN | CFO | STO |
|---|---|---|---|---|---|---|---|---|---|
| Parameter | $(g_{tx},\varphi_{tx})$ | $d^{tx}$ | $\sigma_{tx}^2$ | $(M_\rho,N_\rho)$ | $(g_{rx},\varphi_{rx})$ | $d^{rx}$ | $\sigma_{rx}^2$ | $f_o$ | $(I_1,I_2)$ |
| Scenario 0 | (0 dB,0°) | 0 dB | 0 | (0,0) | (0 dB,0°) | 0 dB | 0 | 0 | (0,0) |
| Scenario 1 | (0.3 dB,1°) | 0.7 dB | 0.1 | (2,2) | (0.3 dB,1°) | 0.7 dB | 0.1 | 15 kHz | (4167,4567) |
| Scenario 2 | (0.8 dB,2°) | 1.2 dB | 0.8 | (3,4) | (0.8 dB,2°) | 1.2 dB | 0.8 | 20 kHz | (4272,4572) |
| Scenario 3 | (1.3 dB,3°) | 1.8 dB | 1.6 | (4,6) | (1.3 dB,3°) | 1.8 dB | 1.6 | 25 kHz | (4388,4588) |
| Scenario 4 | (2 dB,4°) | 2.5 dB | 3 | (5,8) | (2 dB,4°) | 2.5 dB | 3 | 30 kHz | (4495,4595) |

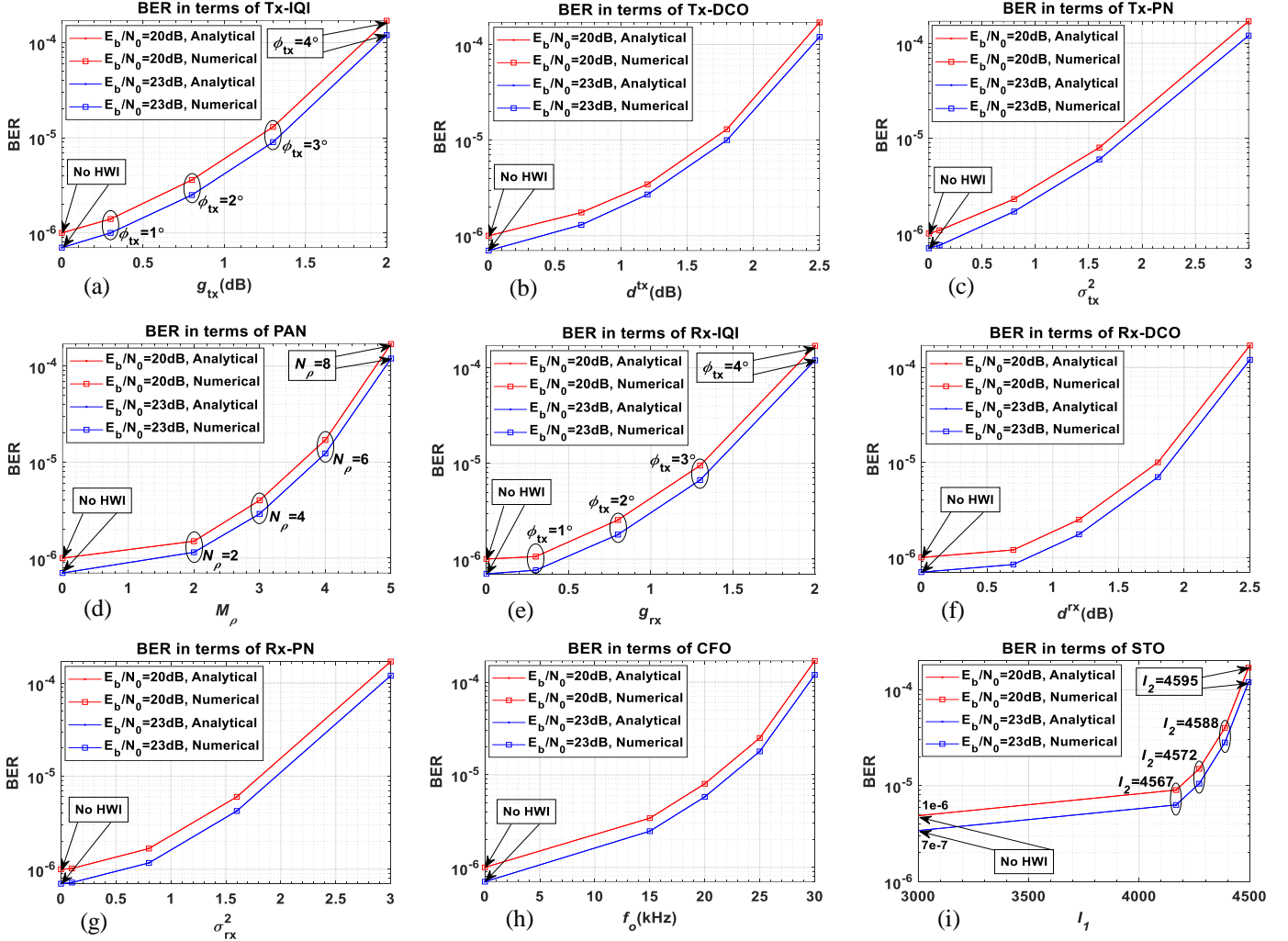

Fig. 5. Analytical and numerical BER performance of coded 4-QAM OTSM with ML detector at mobility speed of 480 kph in terms of: (a) Tx-IQI, (b) Tx-DCO, (c) Tx-PN, (d) PAN, (e) Rx-IQI, (f) Rx-DCO, (g) Rx-PN, (h) CFO, and (i) STO.

In Fig. 5(a), the BER is depicted in terms of Tx-IQI, i.e., the gain and phase mismatches pointed in Table III as $(g_{tx},\varphi_{tx})$, while the other HWI parameters have been fixed at the values listed in the last row of Table III named Scenario 4. It is seen that by increasing the gain and phase mismatches 0.5 dB and 2 times, i.e., moving from $(g_{tx},\varphi_{tx})=(0.3\text{dB},1°)$ to $(g_{tx},\varphi_{tx})=$ (0.8dB,2°), the BER increases 2.32 times and 2.27 times when the SNR is fixed at 20 dB and 23 dB, respectively.

Fig. 5(b) shows the BER in terms of the values of $d^{tx}$ listed in Table III as Tx-DCO. It can be seen that by increasing the Tx-DCO from 0.7 dB to 1.2 dB, the BER increases 1.97 times and 1.95 times at fixed SNR of 20 dB and 23 dB, respectively.

Fig. 5(c) depicts the BER in terms of the variance of Tx-PN given as $\sigma_{tx}^2$ in Table III. As seen, the BER becomes worse 3.47 times and 3.42 times when the Tx-PN variance increases from 0.8 to 1.6 at fixed SNR of 20 dB and 23 dB, respectively.

Fig. 5(d) exhibits the BER as a function of PAN parameters namely the non-linear memory depth and non-linearity order indicated in Table III as $(M_\rho, N_\rho)$. To simulate the effect of PAN, the values of $M_\rho$ and $N_\rho$ are set in a manner that PAN will not be removed from the transmitted signal as given in (12). It is observed that when $M_\rho$ and $N_\rho$ increase 2 times and 3 times, the BER deteriorates 11.33 times and 10.43 times, at fixed SNR of 20 dB and 23 dB, respectively.

In Fig. 5(e), we evaluate the BER in terms of the gain and phase mismatches of Rx-IQI represented in Table III as $(g_{rx}, \varphi_{rx})$. It can be concluded that when Rx-IQI increases 0.5 dB and 2 times, i.e., moving from $(g_{rx}, \varphi_{rx}) = (0.3\text{dB}, 1°)$ to $(g_{rx}, \varphi_{rx}) = (0.8\text{dB}, 2°)$ and other HWIs are fixed to their values in Scenario 4, the BER is raised 2.41 times and 2.35 times at fixed SNR of 20 dB and 23 dB, respectively.

In Fig. 5(f), we show the BER in terms of Rx-DCO denoted as $d^{rx}$ in Table III. It is observed that when DCO increases 0.7 dB from 1.2, the BER increases 2.08 times and 2.07 times at fixed SNR of 20 dB and 23 dB, respectively.

Fig. 5(g) illustrates the BER in terms of the variance of Rx-PN given as $\sigma_{rx}^2$ in Table III. It is seen that the BER becomes worse 3.59 times and 3.55 times when $\sigma_{rx}^2$ increases from 0.8 to 1.6 at fixed SNR of 20 dB and 23 dB, respectively.

From the comparison of Figs. 5(a), 5(b), and 5(c) with Figs. 5(e), 5(f), and 5(g), respectively, and according to the mentioned results, one can conclude the following remarks.

*Remark 7 (The Variation of BER in Terms of Tx-IQI, Tx-DCO, and Tx-PN Are Worse Than Those of Rx-IQI, Rx-DCO, and Rx-PN)*: This is because of two main reasons. First, since in obtaining the variation of BER in terms of Tx-HWIs, Rx-HWIs are set to their harshest values given in Table III through Scenario 4, this keeps the noise power at a high level. Second, in obtaining the variation of BER in terms of the Rx-HWIs, the Tx-HWIs are set on their harshest values given in Scenario 4; however, the values of the related Rx-HWI parameter changes from its smallest value to the largest one, i.e., the noise power slowly increases from its smallest level to its largest one when the BER values are obtained at fixed SNR. Therefore, the BER values for Rx-HWIs are obtained lower than those of Tx-HWIs.

*Remark 8 (The Deterioration of BER Due to the Rx-IQI, Rx-DCO, and Rx-PN Are Larger than Those of Tx-IQI, Tx-DCO, and Tx-PN, Respectively)*: This is because, the Tx-HWIs do not affect on the noise power; however, the power of noise is amplified by the Rx-IQI, Rx-PN, and CFO as analytically discussed in Remark 3, proofed in Theorem 1, and accordingly given in (46). Thus, since an increment in Tx-HWI parameters only affects on the power of signal and an increment in Rx-HWI parameters affects both on the power of signal and noise at the same time, the BER increase due to the Rx-HWIs is always larger than that of Tx-HWIs.

In Fig. 5(h), we provide the variation of BER in terms of CFO given in Table III as $f_o$. The BER increases 2.35 times and 2.34 times when the CFO rises from 15 kHz to 20 kHz at fixed SNR of 20 dB and 23 dB, respectively.

Finally, in Fig. 5(i), we present the BER in terms of the STO parameters given in Table III as $(I_1, I_2)$. Since in this simulation study we aim to evaluate the STO effect, the values of $I_1$ and $I_2$, base-point of interpolant, and fractional interval vectors of interpolating filter given in (17), are set in a fashion that the STO will not be removed. As expected, by increasing the values of $I_1$ and $I_2$ while the less the difference between $I_1$ and $I_2$, the worse the BER performance becomes.

Regarding the effect of SNR increment on the possible BER improvement, it is observed from all mentioned figures that by increasing SNR from 20 dB to 23 dB, the BER improvement is inadequately in the order of 0.01. This is because, in the presence of multiple HWIs and imperfect CSI, increasing the SNR is not a sufficient solution and employing HWIC methods are necessary to provide a more reliable link.

*B. Comparison the Performance of OTSM with OFDM and OTFS under HWIs for Different QAM Orders*

In this simulation example we evaluate the performance of OTSM system by considering all HWIs listed in Scenario 4 of Table III for mobility speed of 480 kph with varying QAM orders. We further evaluate the performance of OFDM and OTFS as the benchmarks. Fig. 6(a) illustrates the average BER in terms of SNR. It is observed that numerical results are close to the analytical upper bound obtained in (55) and they are the same for SNR values larger than 8 dB. On the other hand, there is a gap between the analytical upper bound and numerical results for all SNR values in the case of OFDM and OTFS. This is because in this paper, to obtain both the numerical and analytical results, we consider all HWIs listed in Table I; however, the analytical BER upper bounds obtained in [13] and [18] for OFDM and OTFS, respectively, do not include all of HWIs as stated in Table I. It is worth mentioning that in the absence of HWIs, as explored in [6], OTSM outperforms OFDM and performs close to OTFS in high mobility DSCs. As a new conclusion in this context, it is observed from Fig. 6(a) that OTSM still outperforms OFDM and offers a similar BER performance to OTFS under HWIs and imperfect CSI. At a target BER of $6 \times 10^{-4}$, compared to the ideal system with no HWIs, OFDM undergoes an SNR loss of 10.5 dB; however, OTSM and OTFS experience an SNR loss of 5.5 dB. Thus, OTSM and OTFS outclass OFDM with an SNR gain of 5 dB. It is worth mentioning that evaluating the performance loss of uncoded OTSM system in the presence of joint effects of multiple HWIs at significant error floor is not meaningful to be compared to uncoded OTSM system including no HWIs. Thus, the LDPC coding has been utilized in this paper to reduce the level of error floor but it is not masked. As illustrated in Fig. 6(a), for SNR values more than 24 dB, the BER floors at $6 \times 10^{-4}$. Thanks to the LDPC coding, a suitable region is obtained to evaluate the performance loss of OTSM under the joint effects of multiple HWIs and imperfect CSI compared to ideal system experiencing perfect CSI with no HWIs. We further note that in our analyses, LDPC is used in the standards, and therefore the inclusion of coding in this paper makes it arguably more practical and with real-world impact.

In Fig. 6(b), we further provide the percentage of SNR loss at a target BER of $6 \times 10^{-4}$ for OFDM, OTSM, and OTFS under HWIs when their performances are compared to the OFDM, OTSM, and OTFS ideal systems experiencing no HWIs, respectively. It is observed that in terms of SNR loss, OTSM and OTFS have almost the same percentage and they outperform OFDM for different QAM orders by 29% and 30%, respectively.

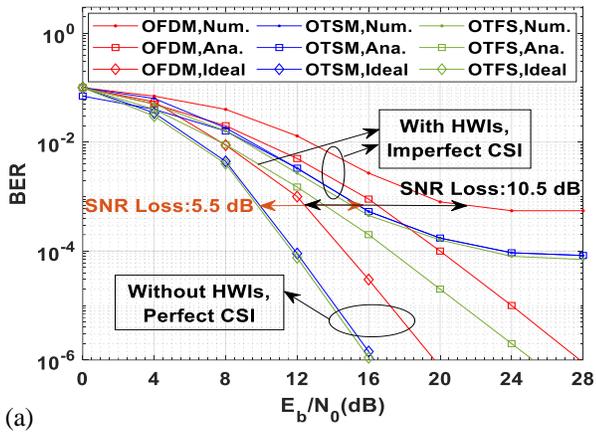

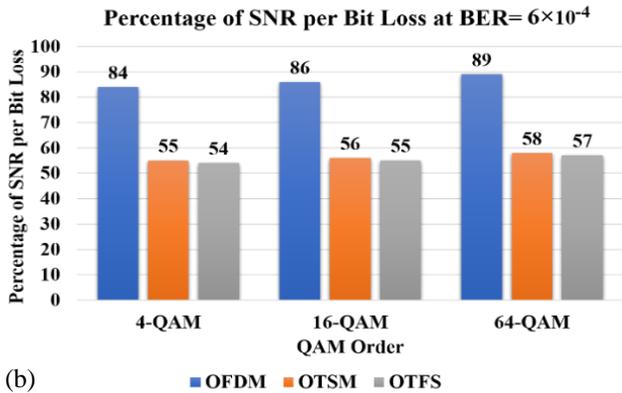

Fig. 6. The BER performance of OTSM compared to OFDM and OTFS under HWIs in Scenario 4 and mobility speed of 480 kph. (a) Comparison the numerical (Num.) and analytical (Ana.) results for 4-QAM. (b) Comparison the percentage of SNR loss for varying QAM orders.

### C. Performance Evaluation for Different Mobility Speeds

In this simulation example, we examine the performance of OTSM system at different mobility speeds by considering the joint effects of all HWIs listed in Table III. Fig. 7(a) shows the BER of OTSM system in terms of SNR under HWI values given in Scenario 4 of Table III. It is seen that at different mobility speeds, numerical results are close to the analytical results and they are the same for SNR values larger than 8 dB, verifying the robustness of OTSM in high mobility DSCs even under the joint effects of multiple HWIs and imperfect CSI. At a target BER of $10^{-4}$ and compared to the ideal system in Scenario 0 of Table III, it is observed that increasing the mobility speed even under the harshest values of HWIs does not change the SNR loss, which is around 11 dB. In Fig. 7(b), we further provide the percentage of SNR loss with varying mobility speeds for different values of HWIs from Scenario 1 to Scenario 4 compared to the ideal system given in Scenario 0. It is observed that by increasing the values of HWIs from Scenario 1 to 4, the percentage of SNR loss increases; however the amount of increment is almost constant for different mobility speeds, i.e., the BER performance does not successively suffer upon increasing the mobility speed even under different values of HWIs.

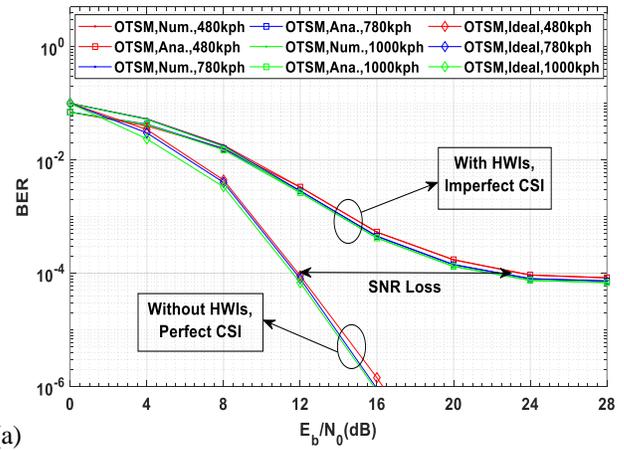

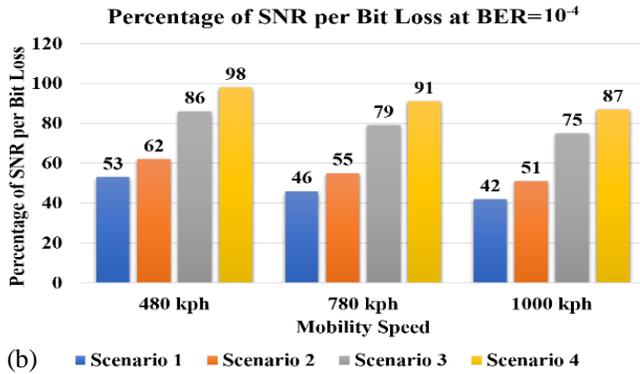

Fig. 7. The BER performance of 4-QAM OTSM under varying HWIs and mobility speeds. (a) Comparison the numerical (Num.) and analytical (Ana.) results for Scenario 4. (b) The Percentage of SNR loss for Scenarios 1 to 4.

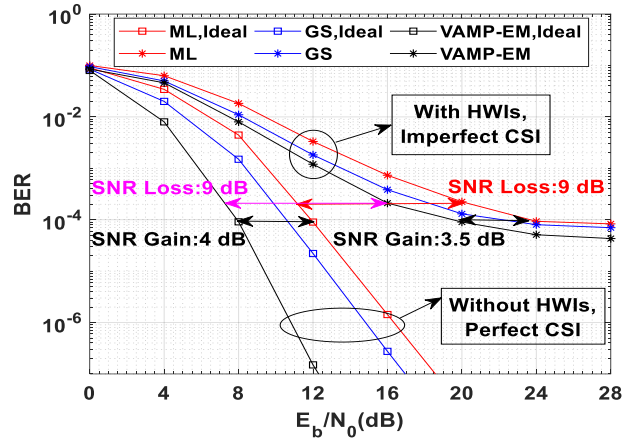

Fig. 8. The BER performance of 4-QAM OTSM at mobility speed of 480 kph for different detectors under HWIs and imperfect CSI based on Scenario 4.

### D. Performance Evaluation for Different Detectors

In this simulation example, we examine the performance of OTSM system by considering the joint effects of all HWIs listed in Table III and imperfect CSI for different detectors. Fig. 8 shows the BER performance in terms of SNR for the ML, matched filtering Gauss Seidel (MFGS) [6], and vector approximate message passing-based expectation maximization (VAMP-EM) [38] detectors. As expected, the VAMP-EM detector outperforms other mentioned detectors even under HWIs and imperfect CSI. This is because of two main reasons making always the detection performance of VAMP-EM better than that of mentioned detectors. First, the a priori information of the modulation scheme is utilized in the VAMP-EM detector in the DS domain [38]. Second, since the VAMP-EM detects the information based on the vector-valued factor graph [38], it will be able to effectively manipulate the ill-conditioned effective channel matrix in the DS domain under imperfect CSI and the joint effects of multiple HWIs.

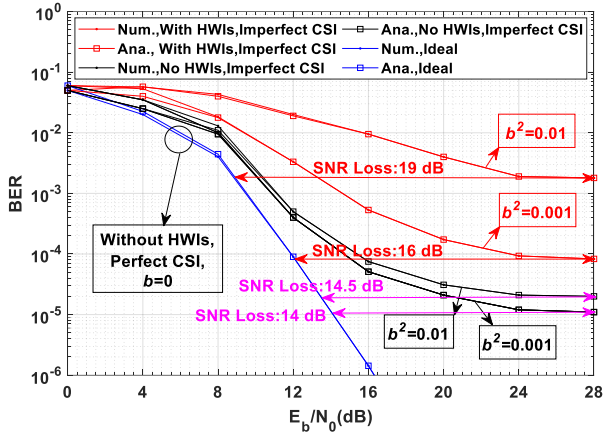

Fig. 9. The BER performance of 4-QAM OTSM at mobility speed of 480 kph for ML detector and different values of imperfectness variance.

*E. Performance Evaluation for Varying CSI Imperfectness*
In this simulation example we assess the performance of OTSM system under HWIs while the ML detector is employed for different orders of imperfect CSI. Fig. 9 shows the BER of 4-QAM OTSM in terms of SNR for varying Variance of CSI imperfectness given as $b^2$ in (48) compared to the case of no HWIs with perfect CSI in which $b$ is equal to zero. It is observed that in the absence of HWIs by employing $b^2 = 0.001$, although the performance floors at a BER of $1.1 \times 10^{-5}$, leading to an SNR loss of 14 dB, the numerical results are close to the analytical upper bound and they are the same for SNR values more than 8 dB. Also, under such conditions, when the imperfectness variance increases 10 times to $b^2 = 0.01$ the BER slowly increases and floors at $2 \times 10^{-5}$. At this BER, the increase of SNR loss due to the increase of imperfectness in the absence of HWIs, is therefore around 4 dB. By considering the HWIs based on Scenario 4 along with the imperfect CSI with a variance of 0.001, the BER floors at $8.3 \times 10^{-5}$, resulting in an SNR loss of 16 dB; however, the numerical and analytical results perfectly match still for SNR values larger than 8 dB. Under such conditions, by increasing the value of imperfectness variance from $b^2 = 0.001$ to $b^2 = 0.01$, the BER significantly increases and floors at $1.8 \times 10^{-3}$, leading to an SNR loss of 19 dB. At this value of BER, the increase of SNR loss due to the increase of imperfectness in the presence of HWIs, is therefore around 10.7 dB. Therefore, under HWIs, the imperfect CSI affects on the system performance much more than in the absence of HWIs. This is because, in the presence of HWIs, the channel estimation error increases due to the amplification of noise by Rx-HWIs as analytically proofed in Theorem 1.

## VII. CONCLUSION

We have derived the discrete-time baseband signal model and the system I/O relations for OTSM transceiver in time, DT, and DS domains by incorporating the joint effects of Tx/Rx IQI, Tx/Rx DCO, Tx/Rx PN, PAN, CFO, and STO. It has been shown that the system I/O relations presented in the previous works on this topic are all obtained as the special cases from our proposed system I/O relations. Thanks to our proposed system I/O relation in the DS domain, we have analytically derived a tight upper bound for the PEP and ABEP in terms of the parameters of mentioned HWIs with imperfect CSI. It has been analytically demonstrated that noise is additive white Gaussian and the effective channel is sparse in the DS domain under joint effects of mentioned HWIs. Meanwhile, Tx/Rx IQI and Tx/Rx DCO introduce self-conjugated and zero sequency interferences, respectively, in the DS domain. We have evaluated the BER performance of LDPC-coded OTSM in mmWave frequency band under HWIs and imperfect CSI through Monte Carlo simulation examples for different QAM orders, mobility speeds, detectors, and CSI imperfectness under different conditions in DSCs. Simulation results have shown that analytical and numerical results make an acceptable match. Thanks to the inherent robustness of OTSM, it is able to perform similar to OTFS and outperforms OFDM in terms of SNR loss by 29% at a target BER of $6 \times 10^{-4}$ under harsh HWIs and imperfect CSI conditions.

## APPENDIX A

*Proof of (34) and Theorem 1:*
The noise vector in the DS domain given in (33) is equal to
$$\widehat{\mathbf{w}} = (\mathbf{I}_M \otimes \mathbf{W}_N)\mathbf{P}^T(\alpha_{rx}\mathbf{R}_{cp}\mathbf{\Theta}_{pn}^{rx}\mathbf{\Theta}_{cfo}\mathbf{w} + \beta_{rx}\mathbf{R}_{cp}\mathbf{\Theta}_{pn}^{rx*}\mathbf{\Theta}_{cfo}^*\mathbf{w}^*), \quad (56)$$
which is a linear combination of Gaussian vectors $\mathbf{w}$ and $\mathbf{w}^*$. Therefore, $\widehat{\mathbf{w}}$ stays Gaussian and since $\mathbf{w}$ is zero-mean, $\widehat{\mathbf{w}}$ becomes zero-mean [39]. Also, by considering the unitary property of WHT, the covariance matrix of $\widehat{\mathbf{w}}$ is given as (57) where $\mathbf{K} = \text{diag}(e^{-j2\omega})$ in which $\omega = 2\pi f_o q + \theta_{pn}^{rx}(q/M\Delta f)$ for $q = 0,\ldots,NM-1$.

$$\text{cov}(\widehat{\mathbf{w}}) = \mathrm{E}(\widehat{\mathbf{w}}.\widehat{\mathbf{w}}^H)$$
$$= \mathrm{E}((\mathbf{I}_M \otimes \mathbf{W}_N)\mathbf{P}^T(\alpha_{rx}\mathbf{R}_{cp}\mathbf{\Theta}_{pn}^{rx}\mathbf{\Theta}_{cfo}\mathbf{w} + \beta_{rx}\mathbf{R}_{cp}\mathbf{\Theta}_{pn}^{rx*}\mathbf{\Theta}_{cfo}^*\mathbf{w}^*)$$
$$.(\alpha_{rx}^*\mathbf{w}^H\mathbf{\Theta}_{cfo}^H\mathbf{\Theta}_{pn}^{rx\,H}\mathbf{R}_{cp}^H + \beta_{rx}^*\mathbf{w}^T\mathbf{\Theta}_{cfo}^T\mathbf{\Theta}_{pn}^{rx\,T}\mathbf{R}_{cp}^H))\mathbf{P}(\mathbf{I}_M \otimes \mathbf{W}_N)^H$$
$$= (\mathbf{I}_M \otimes \mathbf{W}_N)\mathbf{P}^T(|\alpha_{rx}|^2\,\mathbf{R}_{cp}\mathbf{\Theta}_{pn}^{rx}\mathbf{\Theta}_{cfo}\,\underbrace{\mathrm{E}(\mathbf{w}\mathbf{w}^H)}_{\sigma_0^2\mathbf{I}_{NM+l_{\max}}}\mathbf{\Theta}_{cfo}^H\mathbf{\Theta}_{pn}^{rx\,H}\mathbf{R}_{cp}^H$$

$$+ |\beta_{\text{rx}}|^2 \underbrace{\underbrace{\underbrace{\mathbf{R}_{\text{cp}} \mathbf{\Theta}_{\text{pn}}^{\text{rx}\,*} \mathbf{\Theta}_{\text{cfo}}^{*} \underbrace{E(|\mathbf{w}|^2)}_{\sigma_0^2 \mathbf{I}_{NM+l_{\max}}} \mathbf{\Theta}_{\text{cfo}}^{\text{T}} \mathbf{\Theta}_{\text{pn}}^{\text{rx}\,\text{T}} \mathbf{R}_{\text{cp}}^{\text{H}}}_{\sigma_0^2 \mathbf{I}_{NM+l_{\max}}}}_{\sigma_0^2 \mathbf{I}_{NM+l_{\max}}}}_{\sigma_0^2 \mathbf{I}_{NM}}$$

$$+ (\alpha\beta^{*}\mathbf{K})\sigma_0^2 \mathbf{I}_{NM} + (\alpha\beta^{*}\mathbf{K})^{*}\sigma_0^2 \mathbf{I}_{NM} )\mathbf{P}(\mathbf{I}_M \otimes \mathbf{W}_N)^{\text{H}}$$

$$= (\mathbf{I}_M \otimes \mathbf{W}_N)\mathbf{P}^{\text{T}} (|\alpha_{\text{rx}}|^2 + |\beta_{\text{rx}}|^2 + 2\Re(\alpha\beta^{*}\mathbf{K}))\sigma_0^2 \mathbf{I}_{NM} \mathbf{P}(\mathbf{I}_M \otimes \mathbf{W}_N)^{\text{H}}$$

$$= (|\alpha_{\text{rx}}|^2 + |\beta_{\text{rx}}|^2 + 2\Re(\alpha\beta^{*}\mathbf{K}))\sigma_0^2 \mathbf{I}_{NM} (\mathbf{I}_M \otimes \mathbf{W}_N)\mathbf{P}^{\text{T}}\mathbf{P}(\mathbf{I}_M \otimes \mathbf{W}_N)^{\text{H}}$$

$$= (|\alpha_{\text{rx}}|^2 + |\beta_{\text{rx}}|^2 + 2\Re(\alpha\beta^{*}\mathbf{K}))\sigma_0^2 \mathbf{I}_{NM}$$

$$= \frac{1}{2}(1 + \cos(2\omega) + g_{\text{rx}}^2(1 - \cos(2\omega + \varphi_{\text{rx}})))\sigma_0^2 \mathbf{I}_{NM}, \quad (57)$$

Since the Gaussian vector $\widehat{\mathbf{w}}$ is zero-mean and the covariance matrix of $\widehat{\mathbf{w}}$ is diagonal, $\widehat{\mathbf{w}}$ is therefore white Gaussian [39].

APPENDIX B

*Proof of Theorem 2:*
According to the ML criterion given in (47), the probability of transmitting $x_i$ and erroneously detecting $x_j$ conditioned on imperfect CSI vector $\bar{\mathbf{h}}$ is given as

$\Pr(x_i \to x_j | \bar{\mathbf{h}}) =$

$\Pr(\|\mathbf{y}^{\text{HI}} - \mathbf{\Omega}_1(x_i)\mathbf{h} - \mathbf{\Omega}_2(x_i)\mathbf{h}^{*}\|^2 > \|\mathbf{y}^{\text{HI}} - \mathbf{\Omega}_1(x_j)\mathbf{h} - \mathbf{\Omega}_2(x_j)\mathbf{h}^{*}\|^2).$ (58)

Since $\mathbf{\Omega}(\mathbf{x}) = \mathbf{\Omega}_1(\mathbf{x}) + \mathbf{\Omega}_2(\mathbf{x})$, by using $\|\mathbf{A}\|^2 = \text{trace}(\mathbf{A}^{\text{H}}\mathbf{A})$ we obtain $\|\mathbf{\Omega}(\mathbf{x})\mathbf{h}\|^2 = \|\mathbf{\Omega}_1(\mathbf{x})\mathbf{h} + \mathbf{\Omega}_2(\mathbf{x})\mathbf{h}^{*}\|^2$. Then, by utilizing the norm expansion $\|\mathbf{A} - \mathbf{B}\|^2 = \|\mathbf{A}\|^2 + \|\mathbf{B}\|^2 - 2\Re(\mathbf{A}\mathbf{B}^{\text{H}})$, after some simple algebraic manipulations, (58) is simplified to

$\Pr(x_i \to x_j | \bar{\mathbf{h}}) =$

$\Pr(\|\mathbf{\Omega}(x_i)\mathbf{h}\|^2 - \|\mathbf{\Omega}(x_j)\mathbf{h}^{*}\|^2 - 2\Re((\mathbf{y}^{\text{HI}})^{\text{H}}(\mathbf{\Omega}(x_i) - \mathbf{\Omega}(x_j))\mathbf{h}) > 0).$ (59)

Since $\|\mathbf{\Omega}(x_j)\mathbf{h}^{*}\|^2 = \|\mathbf{\Omega}(x_j)\mathbf{h}\|^2$ we obtain

$\Pr(x_i \to x_j | \bar{\mathbf{h}}) =$

$\Pr(-\|(\mathbf{\Omega}(x_i) - \mathbf{\Omega}(x_j))\mathbf{h}\|^2 - 2\Re((\bar{\mathbf{w}})^{\text{H}}(\mathbf{\Omega}(x_i) - \mathbf{\Omega}(x_j))\mathbf{h}) > 0) = \Pr(C > 0).$ (60)

The decision random variable $C$ in (60) follows a Gaussian distribution whose mean and variance are readily obtained as

$$\mu_C = -\|(\mathbf{\Omega}(x_i) - \mathbf{\Omega}(x_j))\mathbf{h}\|^2, \quad (61)$$

$$\sigma_C^2 = 2b^2 \|\mathbf{\Omega}^{\text{H}}(x_i)(\mathbf{\Omega}(x_i) - \mathbf{\Omega}(x_j))\bar{\mathbf{h}}\|^2 + 2\sigma_{\bar{\mathbf{w}}}^2 \|(\mathbf{\Omega}(x_i) - \mathbf{\Omega}(x_j))\bar{\mathbf{h}}\|^2. \quad (62)$$

According to the definition of tail distribution function of Gaussian random variable with mean of $\mu_C$ and variance of $\sigma_C^2$, which is given as $\Pr(C > c_0) = Q((c_0 - \mu_C)/\sigma_C)$ [39], the conditional PEP is obtained as

$\Pr(x_i \to x_j | \bar{\mathbf{h}}) =$

$$Q\left(\frac{\|(\mathbf{\Omega}(x_i) - \mathbf{\Omega}(x_j))\bar{\mathbf{h}}\|^2}{\sqrt{2b^2 \|\mathbf{\Omega}^{\text{H}}(x_i)(\mathbf{\Omega}(x_i) - \mathbf{\Omega}(x_j))\bar{\mathbf{h}}\|^2 + 2\sigma_{\bar{\mathbf{w}}}^2 \|(\mathbf{\Omega}(x_i) - \mathbf{\Omega}(x_j))\bar{\mathbf{h}}\|^2}}\right). \quad (63)$$

Applying the following inequality to (63),

$$\|\mathbf{\Omega}^{\text{H}}(x_i)(\mathbf{\Omega}(x_i) - \mathbf{\Omega}(x_j))\bar{\mathbf{h}}\|^2 \leq \|\mathbf{\Omega}(x_i)\|^2 \|(\mathbf{\Omega}(x_i) - \mathbf{\Omega}(x_j))\bar{\mathbf{h}}\|^2, \quad (64)$$

the conditional PEP can be eventually upper bounded as

$$\Pr(x_i \to x_j | \bar{\mathbf{h}}) \leq Q\left(\sqrt{\frac{\|(\mathbf{\Omega}(x_i) - \mathbf{\Omega}(x_j))\bar{\mathbf{h}}\|^2}{2b^2 \|\mathbf{\Omega}(x_i)\|^2 + 2\sigma_{\bar{\mathbf{w}}}^2}}\right). \quad (65)$$

ACKNOWLEDGEMENT

This work was supported in part by the Natural Science Foundation of China under Grant 62171161; in part by the Shenzhen Science and Technology Program under Grants ZDSYS20210623091808025 (Shenzhen Key Laboratory of Wireless AIoT Communication), KQTD20190929172545139 and GXWD20220817133854003.

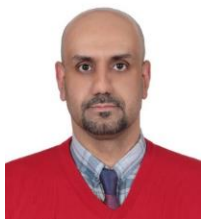

**Abed Doosti-Aref** (Member, IEEE) received his BSc, MSc, and PhD degrees all in telecommunication systems engineering with honors. In 2018, he was awarded by Iran's National Elites Foundation (INEF). In 2020, he was awarded by INEF and Iran National Science Foundation. His current research interests are on physical layer of 6G wireless networks, ISAC in time-frequency, delay-Doppler, delay-sequency, and delay-scale waveform domains including OFDM, OCDM, OSDM, OTFS, OTSM, ODSS, AFDM, ODDM, FM-OFDM, and AI/ML based resource allocation.

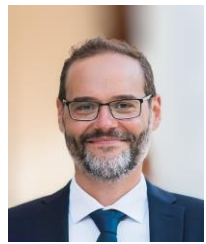

**Christos Masouros** (Fellow, IEEE) is a Professor of signal processing and wireless communications in the Information and Communication Engineering research group, Dept. Electrical and Electronic Engineering, University College London, London, UK. His research interests lie in the field of wireless communications and signal processing with particular focus on green communications, large scale antenna systems, ISAC, interference mitigation techniques for MIMO, and multicarrier communications beyond OFDM.

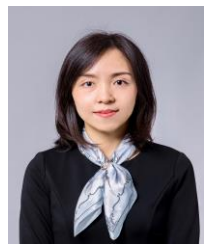

**Xu Zhu** (Senior Member, IEEE) is a Professor of the School of Electronics and Information Engineering, Harbin Institute of Technology, Shenzhen, China. Her research interests lie in the field of wireless communications and signal processing with particular focus on MIMO, channel estimation, ultra reliable low latency communication, resource allocation, interference management, and green communications.

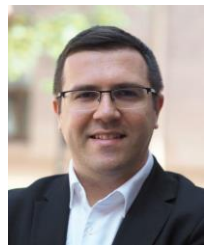

**Ertugrul Basar** (Fellow, IEEE) is an Associate Professor with the Department of Electrical and Electronics Engineering, Koç University, Istanbul, Turkey. His primary research interests include communication theory and systems, reconfigurable intelligent surfaces, index modulation, waveform design, and signal processing for beyond 5G and 6G wireless networks.

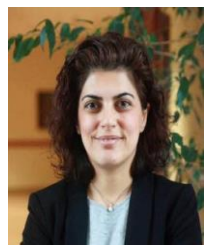

**Sinem Coleri** (Fellow, IEEE) is a Professor and the Chair of the Department of Electrical and Electronics Engineering at Koc University, Istanbul, Turkey. Her research interests are on 6G wireless communications and networking, machine learning for wireless networks, machine-to-machine communications, wireless networked control systems, and vehicular networks.

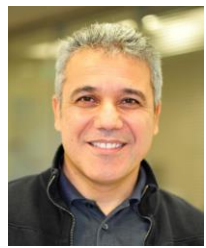

**Huseyin Arslan** (Fellow, IEEE) is a Professor and the Dean of the School of Engineering and Natural Sciences at Istanbul Medipol University, Istanbul, Turkey. His current research interests are on 6G and beyond radio access technologies with particular focus on physical layer security, interference management, cognitive radio, multi-carrier wireless technologies beyond OFDM, dynamic spectrum access, co-existence issues, non-terrestrial communications, and ISAC.